%% file: paper.tex
\pdfoutput=1

\documentclass[longauth]{aa}
\usepackage{graphicx}
\usepackage{gensymb}
\usepackage{txfonts}
\usepackage[flushleft]{threeparttable}
\renewcommand{\arraystretch}{1.1}
\usepackage[hidelinks,colorlinks=true,linkcolor=blue,citecolor=blue]{hyperref}
\usepackage{longtable,tabu}
\usepackage{lipsum}
\usepackage{adjustbox}
\usepackage{stfloats}
\usepackage{scrextend}
\usepackage{subcaption}
\usepackage{amssymb}
\usepackage{caption}
\usepackage{amsmath}
\usepackage{amssymb}
\usepackage{mathrsfs}
\usepackage{multicol}
\usepackage{supertabular}

\newcommand{\feh}{\mbox{[Fe/H]}\xspace}
\newcommand{\teff}{\ensuremath{T_{\rm eff}}}

\newcommand{\kms}{\mbox{km\,s$^{-1}$}\xspace}
\newcommand{\ms}{\mbox{m\,s$^{-1}$}\xspace}

\newcommand{\mjup}{\mbox{$\mathrm{M_{\rm Jup}}$}\xspace}
\newcommand{\rjup}{\mbox{$\mathrm{R_{\rm Jup}}$}\xspace}

\newcommand{\me}{\mbox{$\mathrm{M_{\rm \oplus}}$}\xspace}
\newcommand{\re}{\mbox{$\mathrm{R_{\rm \oplus}}$}\xspace}
\newcommand{\mstar}{\mbox{$M_{*}$}\xspace}
\newcommand{\rstar}{\mbox{$R_{*}$}\xspace}
\newcommand{\densstar}{\mbox{$\rho_*$}\xspace}

\newcommand{\msol}{\mbox{$\mathrm{M_\odot}$}\xspace}

\newcommand{\densjup}{\mbox{$\mathrm{\rho_{Jup}}$}\xspace}

\providecommand{\bjdtdb}{\ensuremath{\rm {BJD_{TDB}}}}
\providecommand{\fave}{\langle F \rangle}

\newcommand{\CONAN}{{\sc \tt CONAN}\xspace}
\newcommand{\celerite}{{\sc \tt celerite}\xspace}

\newcommand{\george}{{\sc \tt george}\xspace}

\newcommand{\ldcu}{{\sc \tt LDCU}\xspace}

\AtBeginDocument{}

\begin{document} 

\title{%
  Three Saturn-mass planets transiting F-type stars revealed with TESS and HARPS\thanks{The photometric and radial velocity data in this work are only available in electronic form at the CDS via anonymous ftp to cdsarc.u-strasbg.fr (130.79.128.5) or via \url{http://cdsarc.u-strasbg.fr/viz-bin/qcat?J/A+A/}}}
\subtitle{TOI-615b, TOI-622b, and TOI-2641b}
\titlerunning{}
\authorrunning{A. Psaridi, et al.}
    \author{Angelica Psaridi
    \inst{\ref{inst-geneva}}\thanks{\email{angeliki.psaridi@unige.ch}}
    \and  Fran\c{c}ois Bouchy \inst{\ref{inst-geneva}}
     \and  Monika Lendl \inst{\ref{inst-geneva}}
     \and Babatunde Akinsanmi\inst{\ref{inst-geneva}}
     \and Keivan G. Stassun \inst{\ref{inst-vandy}}
     \and Barry Smalley \inst{\ref{inst-keele}}
    \and David J. Armstrong\inst{\ref{inst-warwick1},\ref{inst-warwick2}}
     \and Saburo Howard \inst{\ref{inst-cotedazur}}
    \and Sol{\`e}ne Ulmer-Moll\inst{\ref{inst-geneva},\ref{inst-bern}}
   \and   Nolan Grieves\inst{\ref{inst-geneva}}
   \and   Khalid Barkaoui\inst{\ref{inst-liege},\ref{inst-cambridge},\ref{inst-tenerife}}
    \and  Joseph E. Rodriguez\inst{\ref{inst-michigan}}
     \and Edward M. Bryant\inst{\ref{inst-Mullard}}
    \and  Olga Su\'{a}rez\inst{\ref{inst-cotedazur}}
    \and  Tristan Guillot \inst{\ref{inst-cotedazur}}
     \and Phil Evans \inst{\ref{inst-elsauce}}
    \and  Omar Attia \inst{\ref{inst-geneva}}
     \and Robert A. Wittenmyer \inst{\ref{inst-queensland}}
    \and  Samuel W.\ Yee \inst{\ref{inst-princeton}}
    \and  Karen A.\ Collins \inst{\ref{inst-cambridge1}}
    \and  George Zhou \inst{\ref{inst-queensland}}
     \and Franck Galland, \inst{\ref{inst-geneva},\ref{inst-grenoble}}
     \and L\'{e}na Parc \inst{\ref{inst-geneva}}
     \and Stéphane Udry \inst{\ref{inst-geneva}}
     \and Pedro Figueira \inst{\ref{inst-geneva},\ref{inst-porto}}
    \and Carl Ziegler\inst{\ref{inst-austinstate}}
    \and Christoph Mordasini\ \inst{\ref{inst-bern}}
    \and Joshua N.\ Winn \inst{\ref{inst-princeton}}
    \and Sara Seager \inst{\ref{inst-cambridge},\ref{inst-kavli},\ref{inst-Astronautics}}
    \and Jon~M.~Jenkins \inst{\ref{inst-NASAAmes}}
    \and  Joseph~D.~Twicken \inst{\ref{inst-NASAAmes},\ref{inst-SETI}}
    \and Rafael Brahm  \inst{\ref{inst-Adolfo},\ref{inst-Millennium},\ref{inst-DOF}}
    \and Mat\'{i}as I. Jones, \inst{\ref{inst-ESOCHILE}}    
     \and Lyu Abe, \inst{\ref{inst-cotedazur}}
     \and Brett Addison \inst{\ref{inst-queensland},\ref{inst-Swinburne}}
    \and  C\'{e}sar Brice\~{n}o \inst{\ref{inst-tololo}}
    \and Joshua T. Briegal \inst{\ref{inst-Cavendish}}
    \and  Kevin I.\ Collins \inst{\ref{inst-GeorgeMason}}
     \and Tansu Daylan  \inst{\ref{inst-princeton}}
     \and Phillip Eigm{\"u}ller\inst{\ref{inst-DLR}}
    \and  Gabor Furesz \inst{\ref{inst-kavli}}
    \and  Natalia~M.~Guerrero\inst{\ref{inst-florida},\ref{inst-kavli}}
    \and  Janis Hagelberg \inst{\ref{inst-geneva}}
    \and  Alexis Heitzmann \inst{\ref{inst-queensland}}
    \and Rebekah Hounsell \inst{\ref{inst-Maryland},\ref{inst-NASA}}
    \and  Chelsea X. Huang\inst{\ref{inst-queensland}}
    \and Andreas Krenn \inst{\ref{inst-geneva},\ref{inst-Austrian}}
     \and Nicholas M. Law\inst{\ref{inst-chapelhill}}
     \and Andrew W. Mann\inst{\ref{inst-chapelhill}}
     \and James McCormac \inst{\ref{inst-warwick1},\ref{inst-warwick2}}
     \and Djamel M\'{e}karnia \inst{\ref{inst-cotedazur}}
     \and Dany Mounzer \inst{\ref{inst-geneva}}
     \and Louise D. Nielsen\inst{\ref{inst-eso}}
     \and Ares Osborn \inst{\ref{inst-warwick1},\ref{inst-warwick2}}
     \and Yared Reinarz\inst{\ref{inst-Millennium}}
     \and Ramotholo R. Sefako \inst{\ref{inst-capetown}}
    \and  Michal Steiner\inst{\ref{inst-geneva}}
    \and  Paul A. Str{\o}m\inst{\ref{inst-warwick1},\ref{inst-warwick2}}
     \and Amaury H.M.J. Triaud \inst{\ref{inst-Birmingham}}
     \and Roland Vanderspek\inst{\ref{inst-cambridge},\ref{inst-kavli}}
     \and Leonardo Vanzi \inst{\ref{inst-Pontificia}}
    \and  Jose I. Vines\inst{\ref{inst-UnideChile}}
    \and  Christopher A. Watson \inst{\ref{inst-Belfast}}
    \and  Duncan J. Wright \inst{\ref{inst-queensland}}
    \and  Abner Zapata\inst{\ref{inst-Pontificia}}}
 \institute{
    Observatoire de Gen{\`e}ve, Universit{\'e} de Gen{\`e}ve, Chemin Pegasi 51, 1290 Versoix, Switzerland \label{inst-geneva}
  \and
   Vanderbilt University, Department of Physics \& Astronomy, 6301 Stevenson Center Lane, Nashville, TN 37235, USA \label{inst-vandy}
  \and
    Astrophysics Group, Keele University, Staffordshire ST5 5BG, UK \label{inst-keele}
  \and
    Department of Physics, University of Warwick, Gibbet Hill Road, Coventry CV4 7AL, UK, USA \label{inst-warwick1}
  \and
    Centre for Exoplanets and Habitability, University of Warwick, Gibbet Hill Road, Coventry CV4 7AL, UK \label{inst-warwick2}
    \and 
    Universit\'e C\^ote d'Azur, Observatoire de la C\^ote d'Azur, CNRS, Laboratoire Lagrange, Bd de l'Observatoire, CS 34229, 06304 Nice cedex 4, France \label{inst-cotedazur}
  \and
  Physikalisches Institut, University of Bern, Gesellschaftsstrasse 6, 3012 Bern, Switzerland \label{inst-bern}
  \and
   Astrobiology Research Unit, Universit\'e de Li\`ege, 19C All\'ee du 6 Ao\^ut, 4000 Li\`ege, Belgium \label{inst-liege}  
  \and
    Department of Earth, Atmospheric and Planetary Science, Massachusetts Institute of Technology, 77 Massachusetts Avenue, Cambridge, MA 02139, USA \label{inst-cambridge}  
  \and
    Instituto de Astrof\'isica de Canarias (IAC), Calle V\'ia L\'actea s/n, 38200, La Laguna, Tenerife, Spain \label{inst-tenerife}   
  \and
Center for Data Intensive and Time Domain Astronomy, Department of Physics and Astronomy, Michigan State University, East Lansing, MI 48824, USA \label{inst-michigan}
\and 
  Mullard Space Science Laboratory, University College London, Holmbury St Mary, Dorking, Surrey, RH5 6NT, UK \label{inst-Mullard}  
  \and
 El Sauce Observatory, Coquimbo Province, Chile \label{inst-elsauce}
  \and
    University of Southern Queensland, Centre for Astrophysics, West Street, Toowoomba, QLD 4350, Australia \label{inst-queensland}
  \and
    Department of Astrophysical Sciences, Princeton University, Princeton, NJ 08544, USA \label{inst-princeton}
  \and
    Center for Astrophysics \textbar \ Harvard \& Smithsonian, 60 Garden Street, Cambridge, MA 02138, USA \label{inst-cambridge1}
  \and
 University of Grenoble Alpes, CNRS, IPAG, F-38000 Grenoble, France \label{inst-grenoble}
  \and
Instituto de Astrof\'{i}sica e Ci\^encias do Espa\c{c}o, Universidade do Porto, CauP, Rua das Estrelas, 4150-762 Porto, Portugal \label{inst-porto}
  \and
  Department of Physics, Engineering and Astronomy, Stephen F. Austin State University, 1936 North St, Nacogdoches, TX 75962, USA \label{inst-austinstate}
  \and
  Department of Physics and Kavli Institute for Astrophysics and Space Research, Massachusetts Institute of Technology, Cambridge, MA 02139, USA \label{inst-kavli}
  \and
  Department of Aeronautics and Astronautics, Massachusetts Institute of Technology, Cambridge, MA 02139, USA \label{inst-Astronautics}
  \and
SETI Institute, Mountain View, CA 94043, USA \label{inst-SETI}
\and
Facultad de Ingeniera y Ciencias, Universidad Adolfo Ib\'{a}\~{n}ez, Av. Diagonal las Torres 2640, Pe\~{n}alol\'{e}n, Santiago, Chile \label{inst-Adolfo}
\and
Millennium Institute for Astrophysics, Chile\label{inst-Millennium}
\and
Data Observatory Foundation, Chile\label{inst-DOF}
\and
European Southern Observatory, Alonso de C\'ordova 3107, Vitacura, Casilla, 19001, Santiago, Chile \label{inst-ESOCHILE}
  \and
  NASA Ames Research Center, Moffett Field, CA 94035, USA \label{inst-NASAAmes} 
  \and
  Swinburne University of Technology, Centre for Astrophysics and Supercomputing, John Street, Hawthorn, VIC 3122, Australia\label{inst-Swinburne} 
  \and
    Cerro Tololo Inter-American Observatory/NSF's NOIRLab, Casilla 603, La Serena, Chile \label{inst-tololo}
  \and
   Astrophysics Group, Cavendish Laboratory, J.J. Thomson Avenue, Cambridge, CB3 0HE, UK \label{inst-Cavendish}
  \and
    George Mason University, 4400 University Drive, Fairfax, VA, 22030 USA \label{inst-GeorgeMason}  
  \and
   Department of Extrasolar Planets and Atmospheres, Institute of Planetary Research, German Aerospace Center (DLR), Rutherfordstraße 2, D-12489 Berlin, Germany \label{inst-DLR}
  \and
  Bryant Space Science Center, Department of Astronomy, University of Florida, Gainesville, FL 32611, USA \label{inst-florida}
  \and
  University of Maryland, Baltimore County, Baltimore, MD 21250, USA \label{inst-Maryland}
  \and
  NASA Goddard Space Flight Center, 8800 Greenbelt Rd, Greenbelt, MD 20771, USA \label{inst-NASA}
  \and
    Space Research Institute, Austrian Academy of Sciences, Schmiedlstraße 6, 8042 Graz, Austria \label{inst-Austrian} 
  \and
    Department of Physics and Astronomy, The University of North Carolina at Chapel Hill, Chapel Hill, NC 27599-3255, USA \label{inst-chapelhill}
    \and
 European Southern Observatory, Karl-Schwarzschild-Stra${\ss}$e 2, 85748 Garching bei M{\"u}nchen, Germany \label{inst-eso}
  \and
  Instituto de Astronomía, Universidad Católica del Norte, Angamos 0610, 1270709, Antofagasta, Chile \label{inst-Antofagasta}
  \and
  South African Astronomical Observatory, P.O. Box 9, Observatory, Cape Town 7935, South Africa \label{inst-capetown}
  \and
  School of Physics \& Astronomy, University of Birmingham, Edgbaston, Birmingham B15 2TT, UK \label{inst-Birmingham}
  \and
  Department of Electrical Engineering and Center of Astro Engineering, Pontificia Universidad Católica de Chile, Av. Vicuña Mackenna 4860, Santiago, Chile \label{inst-Pontificia}
\and
 Departamento de Astronomia, Universidad de Chile, Casilla 36-D, Santiago, Chile \label{inst-UnideChile}
\and
 Astrophysics Research Centre, Queen's University Belfast, Belfast BT7 1NN, UK \label{inst-Belfast}
}
   \date{Received March 14, 2023; Accepted May 2, 2023}
   \abstract{While the sample of confirmed exoplanets continues to grow, the population of transiting exoplanets around early-type stars is still limited.  These planets allow us to investigate the planet properties and formation pathways over a wide range of stellar masses and study the impact of high irradiation on hot Jupiters orbiting such stars. We report the discovery of TOI-615b, TOI-622b, and TOI-2641b, three Saturn-mass planets transiting main sequence, F-type stars. The planets were identified by the Transiting Exoplanet Survey Satellite (TESS) and confirmed with complementary ground-based and radial velocity observations. TOI-615b is a highly irradiated ($\sim$1277 $F_{\oplus}$) and bloated Saturn-mass planet (1.69$^{+0.05}_{-0.06}$~\rjup and 0.43$^{+0.09}_{-0.08}$~\mjup) in a 4.66 day orbit transiting a 6850 K star. TOI-622b has a radius of 0.82$^{+0.03}_{-0.03}$~\rjup and a mass of 0.30$^{+0.07}_{-0.08}$~\mjup in a 6.40 day orbit. Despite its high insolation flux ($\sim$600 $F_{\oplus}$), TOI-622b does not show any evidence of radius inflation. TOI-2641b is a 0.39$^{+0.02}_{-0.04}$~\mjup planet in a 4.88 day orbit with a grazing transit (b = 1.04$^{+0.05}_{-0.06 }$) that results in a poorly constrained radius of 1.61$^{+0.46}_{-0.64}$~\rjup. Additionally, TOI-615b is considered attractive for atmospheric studies via transmission spectroscopy with ground-based spectrographs and $\textit{JWST}$. Future atmospheric and spin-orbit alignment observations are essential since they can provide information on the atmospheric composition, formation, and migration of exoplanets across various stellar types.}


   \maketitle
%

\section{Introduction}\label{sec:Introduction}
To date, more than 122 extrasolar planets with well-constrained densities\protect\footnote{$\sigma_{M}/M$ $\leq$ 25$\%$ and $\sigma_{R}/R$ $\leq$8$\%$, resulting in density uncertainty $\leq$ 20$\%$ \label{precision}} that are transiting AF-type, main-sequence stars above the Kraft break (\teff~$\gtrsim$~6200~K, $\lesssim$~F8V, or  \mstar~$\gtrsim$~1.1~\msol; \citealt{kraft}) have been discovered according to the PlanetS exoplanet catalog\protect\footnote{Available on the Data \& Analysis Center for Exoplanets (DACE) platform (\url{https://dace.unige.ch})\label{dacefootnote}} (\citealt{otegi2020}). Obtaining mass measurements of planets orbiting such stars is inherently more difficult than for Solar-type stars. These stars have thin convective envelopes, do not efficiently generate magnetic winds, and often rotate rapidly. Consequently, they exhibit broader and fewer spectral lines due to their rapid rotation and high temperature, complicating high precision radial velocity (RV) measurements.

Hot Jupiters that closely orbit early-type stars are scientifically interesting since their high equilibrium temperatures make them valuable for atmospheric characterization via transmission spectroscopy. Additionally, 20 out of the 35 planets on misaligned orbits ($|\mathrm{\lambda}|$>10$^{\circ}$, with >3$\sigma$ confidence) have been found to orbit stars with effective temperatures above the Kraft break (DACE\textsuperscript{\ref{dacefootnote}}). Their high obliquities, compared to planets orbiting cooler stars (\citealt{winn2010}), can provide insight into the formation of the system, suggesting that the planets were formed ex situ and migrated inward due to interactions with the protoplanetary disk.

NASA's Transiting Exoplanet Survey Satellite (TESS, \citealt{Ricker2015}), in contrast to Kepler (\citealt{Borucki2010}), is focusing on bright targets including hot, early-type stars. TESS has delivered over 350 planetary candidates around stars with effective temperatures above the Kraft break so far - these objects necessitate further spectroscopic and photometric observations to confirm their planetary nature. To date, 25 exoplanets around early-type stars have been confirmed with such follow-up studies:  HD 202772Ab (\citealt{HD202772}), HD 2685b (\citealt{HD2685}), TOI-150b, TOI-163b (\citealt{TOI150TOI163}), TOI-677b (\citealt{jordan:2020}), TOI-892b (\citealt{brahm:2020}), HD 5278b (\citealt{HD5278}), TOI-628b, TOI-640b, TOI-1333b (\citealt{TOI1333TOI628640}), TOI-1431b (\citealt{TOI1431}), TOI-201b (\citealt{toi201}), TOI-2109b (\citealt{TOI2109}), TOI-3362b (\citealt{TOI3362}), TOI-1107b (\citealt{Psaridi2022}), TOI-1516b, TOI-2046b (\citealt{TOI1516TOI2046}), TOI-1842b (\citealt{TOI1842}), TOI-4137b (\citealt{TOI4137}), TOI-5153b (\citealt{TOI5153}), TOI-558b (\citealt{TOI558}), TOI-2152b, TOI-2154b, TOI-2497b (\citealt{Rodriguez2022}), and TOI-2803Ab (\citealt{Yee2023}).

Given the interest in understanding the planet formation process over a wide range of stellar characteristics and especially for massive stars, we began a survey characterizing TESS candidates around a sample of AF main-sequence stars with ground-based follow-up spectroscopy. Our target list consists of relatively bright stars (V $\le$ 12 mag) with effective temperatures above the Kraft break that host giant exoplanet and brown dwarf candidates (R > 5 \re). This survey resulted in the detection of a new massive planet, TOI-1107b, and three brown dwarfs, TOI-629b, TOI-1982b, and TOI-2543b (\citealt{Psaridi2022}). In this work, we report the discovery and confirmation of three new Saturn-mass planets, TOI-615b, TOI-622b, and TOI-2641b, orbiting around F-type stars. The planets were initially detected in the TESS data during the first three years of its mission (Sectors 1-39). In addition, we performed follow-up spectroscopic observations with HARPS, CORALIE and CHIRON, MINERVA-Australis, FIDEOS, and NRES, and ground-based photometry with EulerCam, NGTS, ASTEP, LCOGT, and El Sauce.

The paper is structured as follows: in Section \ref{sec:Observations} we describe the photometric and follow-up spectroscopic observations. Section \ref{sec:analysis} details the stellar characterization and the global analysis. Finally, in Section \ref{sec:Discussion and conclusion} we discuss the planets in context and the
feasibility for future studies.

\section{Observations}\label{sec:Observations}

\input{tables/tessandgroundphotometry}

\subsection{Photometry}\label{sec:photometry}
The transits of the three targets were initially observed and detected in TESS photometry during the first three years of its mission (Sectors 1-39). Afterward, we ruled out nearby eclipsing binaries (NEBs), confirmed the transit depth and refined the ephemeris with follow-up photometry as part of the TESS Follow-up Observing Program (TFOP, \citealt{karen2018}). The summary of the photometric observations can be found in Table~\ref{tab:tessandgroundphotometry}. All the light curves are presented in Figure \ref{fig:Light curves} in the Appendix. 

\subsubsection{TESS photometry}\label{sec:TESS_photometry}
TOI-615 (TIC 190496853) was observed in Sectors 8 and 9 with a 30-minute cadence and Sector 35 with a 2-minute cadence. Nine transits were detected in the Full Frame Images (FFIs) and four transits in the 2-minute cadence light curves with a transit depth of $\sim$ 1.12$\%$ and an average period of 4.66 days. TOI-615 passed all the Science Processing Operations Center (SPOC; \citealt{Jenkins2016}) Data Validation (DV) tests (\citealt{Twicken2018}, \citealt{Li2019}) and the difference image centroiding analysis for Sector 35 that located the transit signature source to within 0.7 $\pm$ 2.5$\arcsec$ of the target star.

TOI-622 (TIC 83092282) was observed in Sectors 8, 9, 34, and 35 at 2-minute cadence where 14 transits were detected with a depth of $\sim$ 0.38$\%$ and a period of 6.4 days. The transit signature passed all DV tests and the difference image centroid offsets for the four available Sectors located the source of the transits to within 1.5 $\pm$ 2.5$\arcsec$ of the target star.

Finally, TOI-2641 (TIC 162802770) was monitored in Sectors 36 and 37 at 2-minute cadence. Nine transits with an orbital period of 4.88 days and a depth of $\sim$ 0.28$\%$ were identified in the data, suggesting a $\sim$ 6.9 \re companion radius. The candidate passed all DV tests and the difference image centroid offsets for Sectors 36-37 located the source of the transits to within 2.1 $\pm$ 2.7$\arcsec$ of the target star. In Section \ref{sec:Global analysis} we show that the impact parameter is greater than one and the planetary radius is much larger than the initial estimate. The high impact parameter and the V-shaped transits suggest that the companion is in a highly grazing transit.

The 2-minute cadence photometry is extracted by the SPOC pipeline which produces the Simple Aperture Photometry (SAP) flux and the Presearch Data Conditioning SAP (PDCSAP; \citealt{Smith2012}, \citealt{Stumpe2012}, \citeyear{Stumpe2014}) flux. The 30-minute photometry was produced by the MIT Quick Look Pipeline (QLP; \citealt{QLP}). The SPOC pipeline transiting planet search component employs an adaptive, noise-compensating matched filter (\citealt{Jenkins2002}, \citeyear{Jenkins2010}, \citeyear{Jenkins2020}). The QLP searches light curves for periodic transits with the box-least-squares algorithm (BLS; \citealt{Kovacs2002}). Alerts for planet candidate TOIs (\citealt{Guerrero2021}) are issued by the TESS Science Office after vetting the SPOC and QLP transit detections.


\subsubsection{Ground-based follow-up photometry}\label{sec:groundphotometry}
We observed a full transit of TOI-615 and TOI-2641 in V-band using the EulerCam instrument (\citealt{Lendl2012}) installed on the 1.2-meter Leonhard Euler Telescope at ESO’s La Silla Observatory. TOI-615 was observed on 2022 February 16 with an exposure time of 90 s and a defocus of 0.12 mm. TOI-2641 was observed on 2022 February 28 with an exposure time of 75 s and a defocus of 0.08 mm. We reduced the data with the standard procedure of bias subtraction and flat correction. For the reduction of the light curves we performed differential aperture photometry with a careful selection of aperture radius and comparison stars that minimize the RMS scatter in the out of transit portion.

TOI-622b was observed with the Next Generation Transit Survey (NGTS; \citealt{NGTS2018}) on 2019 December 22. NGTS is a wide-field photometric survey installed at ESO’s Paranal observatory in Chile that consists of an array of twelve individual robotic 20\,cm telescopes, and by using multiple telescopes to simultaneously observe the same star can achieve high precision photometry for bright stars \citep{smith2020, bryant2020}. Two telescopes were used simultaneously covering the ingress and mid part of the transit with the custom NGTS filter (520-890 nm). A total of 2421 images were collected with an exposure time of 10\,seconds. A custom aperture photometry pipeline, which uses the SEP Python library \citep{BertinArnouts1996, Barbary2016} to perform the source extraction and photometry, was used to reduce the NGTS data. During this reduction, comparison stars which are similar to TOI-622 in brightness, color and CCD position are automatically selected using \textit{Gaia} DR2 (\citealt{GaiaCollaboration2018}).

We observed three full transits of TOI-2641 using the Las Cumbres Observatory Global Telescope (LCOGT; Brown et al. 2013) 1.0-m network node at Cerro Tololo Interamerican Observatory. The telescope is equipped with 4096 $\times$ 4096 SINISTRO Cameras having an image scale of 0.389 arcsec/pixel, resulting in a 26' $\times$ 26' field of view. We used the TESS Transit Finder (TTF), which is a customized version of the Tapir software package (\citealt{tapir}), to schedule our transit observations. The first transit was observed on 2021 May 21 in the Sloan-i' filter with an exposure time of 16s and a photometric aperture radius of 3.5 arcsec (9 pixel). Second and third transits were observed on UTC 2021 May 26 in the Sloan-'i and Sloan-g' and a photometric aperture radius of 4.3 arcsec (11 pixel).
The science images were calibrated (bias and dark subtraction and flat division) by the standard LCOGT BANZAI pipeline (\citealt{McCully:2018}), and photometric measurements were extracted using AstroImageJ (\citealt{Collins:2017}). A transit-like event consistent with the SPOC-reported event is detected in all three light curves. The follow-up apertures exclude flux from all known Gaia DR3 and TIC version 8 neighbors, except TIC 940863891, which is $3\farcs4$ northwest of TOI-2641. However, it is fainter than TOI-2641 by 6.77 magnitudes, so is nominally too faint to be capable of producing the $\sim2.7$ ppt event in the SPOC photometric aperture. Nevertheless, we also extracted a light curve from 2021 May 21 observation using a smaller $2\farcs0$ target aperture, which excluded most of the flux from the 3.4" neighbor, and also detect a consistent event on-target although with lower photometric precision.

We observed a partial transit of TOI-615 on 2021 March 18, with the Antarctica Search for Transiting ExoPlanets (ASTEP) program \citep{guillot2015, mekarnia2016}. The 0.4\,m robotic telescope, located at  Concordia station (Dome\,C, Antarctica) , is equipped with an FLI Proline  KAF-16801E, $4096\times 4096$ pixel  CCD camera, observing in a roughly Rc band-pass with a  1$^{o}\times 1^{o}$ field of view. Due to the extremely low data transmission rate at the Concordia station, the data are processed on-site using an automated IDL-based pipeline described in \citep{abe2013}.   The raw light curves of about 1,000 stars are transferred to Europe on a server in Rome, Italy, and are then available for deeper analysis. These data files contain each star’s flux computed through $10$ fixed circular aperture radii so that optimal light curves can be extracted. For TOI-615 a 8.4 arcsec radius aperture gave the best result.


An egress of TOI-615 was observed on 2020 February 07 in Johnson B-band using the Evans 0.36m telescope at El Sauce Observatory in Coquimbo Province, Chile. The telescope is equipped with a STT 1603-3 CCD camera with 1536 $\times$ 1024 pixels binned 2 $\times$ 2 in-camera resulting in an image scale of 1.47" /pixel. The photometric data were obtained from 311 $\times$ 75 sec exposures, after standard calibration, using a circular 7.4" aperture in AstroImageJ (\citealt{Collins:2017}).

\subsection{Spectroscopy}

\subsubsection{Reconnaissance spectroscopy}\label{sec:Reconnaissance}
We collected high resolution, low precision spectra of TOI-615 and TOI-622 that ruled out false positive scenarios, such as blended eclipsing binaries and starspots (\citealt{Queloz2001A}). The reconnaissance measurements are not precise enough to help constrain the planetary parameters, and so we do not include them in the global analysis.

TOI-615 was observed 33 times between 2019 December 06 and 2021 April 30  with CHIRON spectrograph (see Section \ref{sec:Confirmation spectroscopy}) with an exposure time of 1000 s which translated in spectra with a signal-to-noise ratio per resolution element (S/N) between 29 and 77. The RV RMS is 155 \ms and the mean measurement error is 140 \ms, which is $\sim$3.3 times the RV semi-amplitude of TOI-615b.

We obtained 56 reconnaissance spectra for TOI-622 between 2019 April 16 and 2021 January 12 using the MINERVA-Australis telescope array (\citealt{addison19}), located at Mt. Kent Observatory, Australia. Minerva-Australis is an array of four identical 0.7 m telescopes linked via fiber feeds to a single KiwiSpec echelle spectrograph at a spectral resolving power of $R\sim$80,000 over the wavelength region of 5000-6300\AA. Radial velocities for the observations are derived for each telescope by cross-correlation, where the template being matched is the mean spectrum of each telescope. Each epoch consists of 30-60 minute exposures from up to four individual telescopes; the concurrent radial velocities from multiple telescopes are 
binned into a single point after accounting for the offsets between 
telescopes. The RV RMS is 106 \ms and the average measurement error is 151 \ms corresponding to $\sim$5.4 times the semi-amplitude of TOI-622b. We collected 12 spectra using Fiber Dual Echelle Optical Spectrograph (FIDEOS; \citealt{Vanzi2018}) spectrograph mounted on the ESO 1.0 m telescope at La Silla Observatory. The observations were obtained between 2019 May 15 and 2020 February 22 with an exposure time of 1200~s resulting in a S/N between 56 and 94. The RV RMS is 42 \ms and the mean measurement error is 40 \ms ($\sim$1.4 times the RV semi-amplitude of TOI-622b). Additionally, we obtained 10 observations of TOI-622 from 2019 October 08 to 2019 November 02 with the Network of Robotic Echelle Spectrographs (NRES; \citealt{Siverd2018}) on the Las Cumbres Observatory (LCOGT; \citealt{Brown:2013}) with an exposure time of 480 s and S/N between 11 and 49. The RV RMS is 89 \ms and the mean measurement error is 81 \ms, which is $\sim$2.9 times the RV semi-amplitude of TOI-622b. 

\subsubsection{Confirmation spectroscopy}\label{sec:Confirmation spectroscopy}
\begin{figure*}
  \centering
  \includegraphics[width=0.32\textwidth]{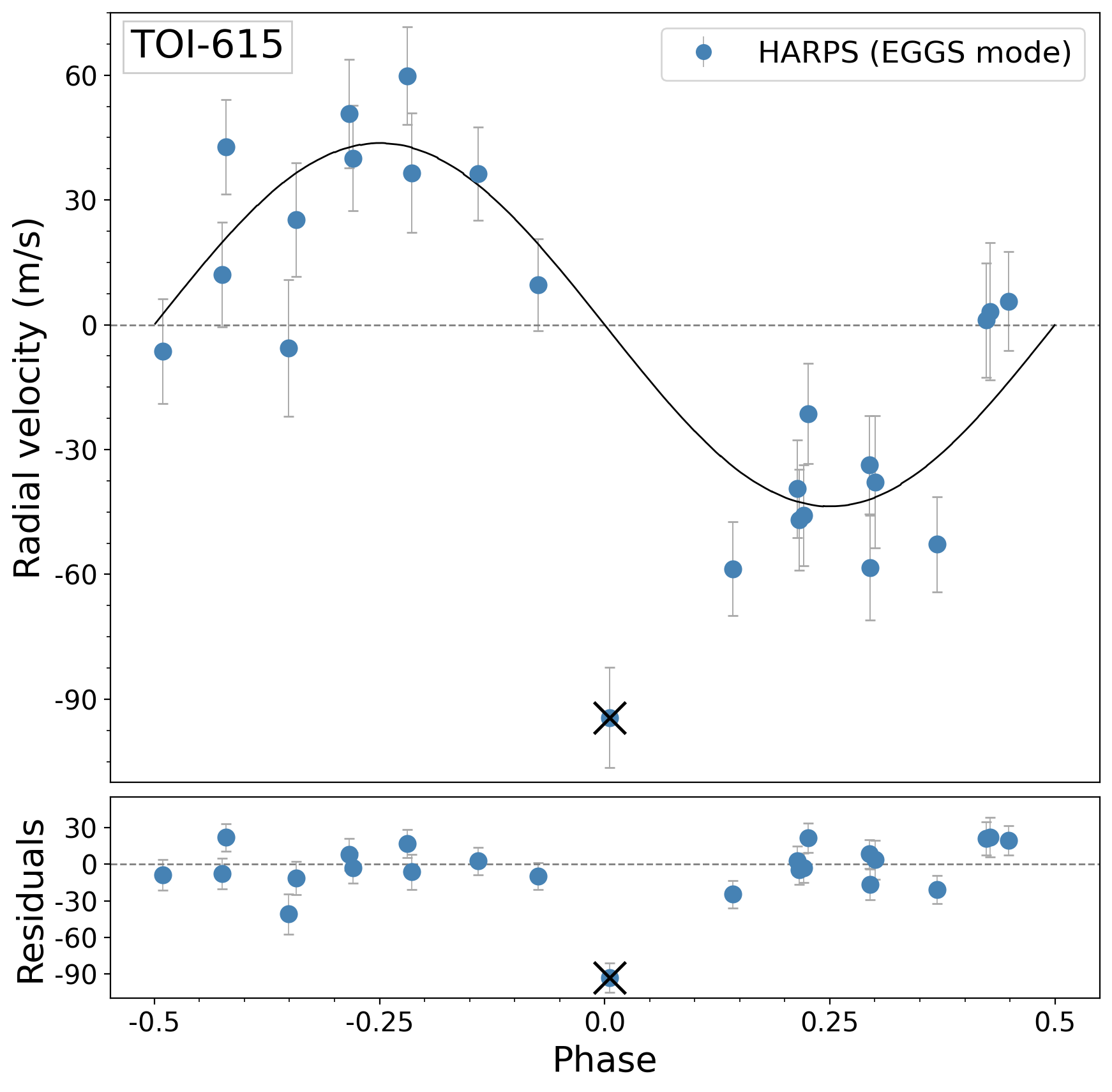}
  \hspace{0.2cm}
  \includegraphics[width=0.32\textwidth]{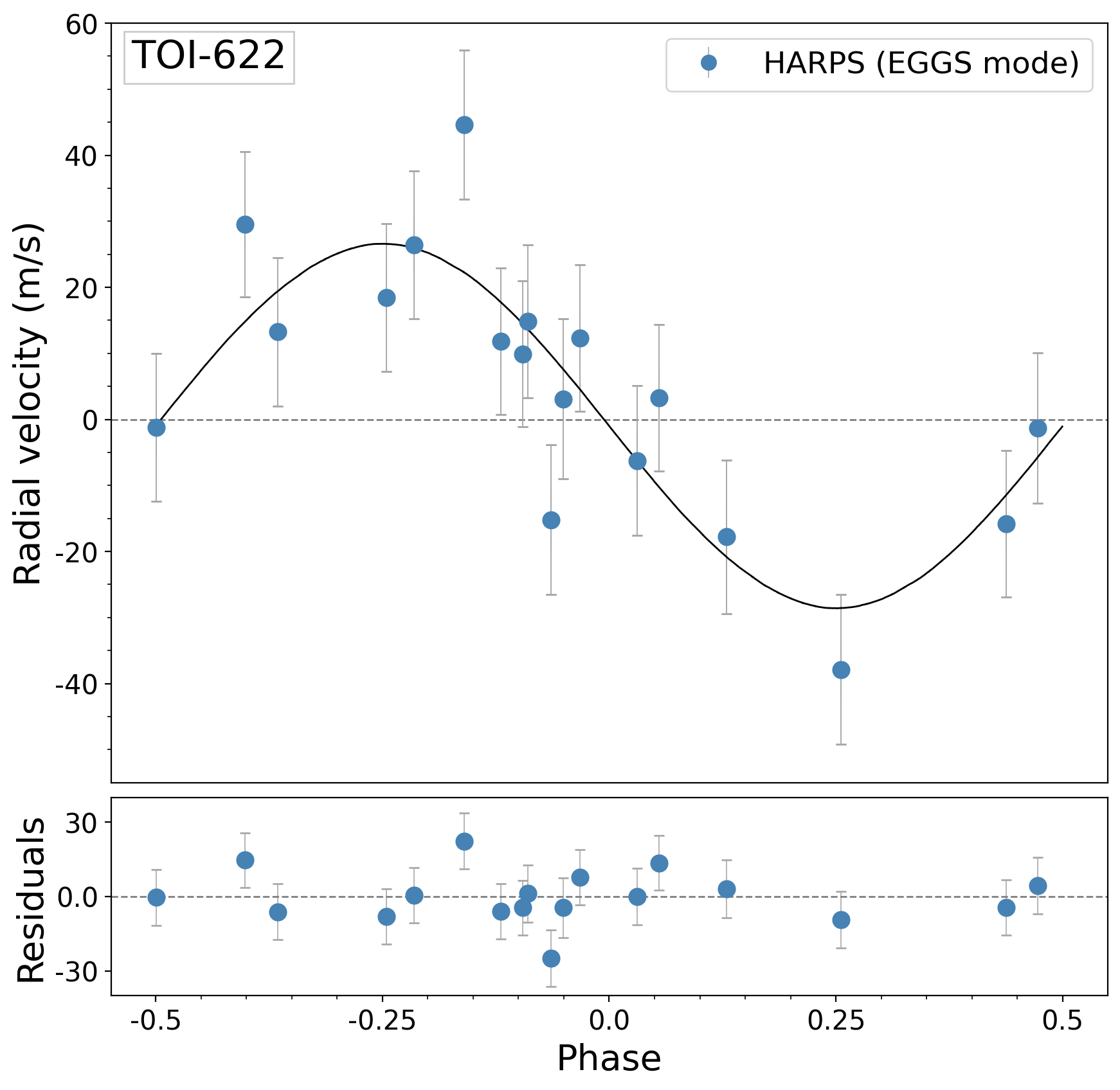}
  \includegraphics[width=0.32\textwidth]{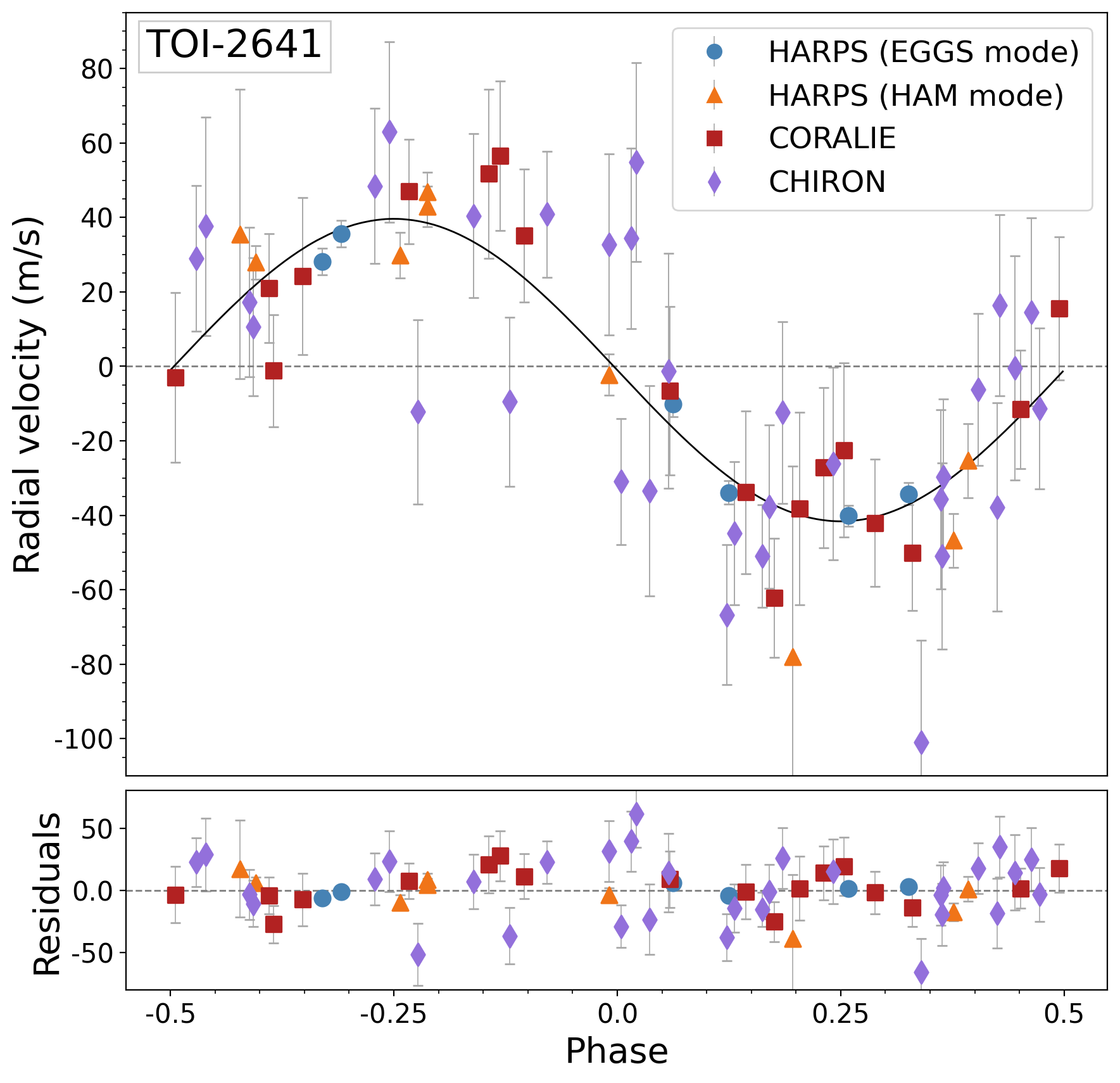}
  \caption{Relative RVs of TOI-615 (\textit{left}), TOI-622 (\textit{middle}), and TOI-2641 (\textit{right}). The different colors indicate different instruments. The black line shows the model fit derived with \CONAN. The residuals of the model fit are shown in the bottom panels. For TOI-615 the crossed-out HARPS data point was collected during transit and is excluded from the analysis.} 
  \label{fig:rvs}
\end{figure*}

Following the reconnaissance observations, we carried out follow-up, higher resolution and more precise spectroscopic observations of the three targets with HARPS, CORALIE, and CHIRON to determine the stellar parameters and obtain precise mass measurements. The phase-folded RVs with the best-fit model from the global analysis can be found in Figure \ref{fig:rvs}.

We collected high-resolution spectra using the High Accuracy Radial Velocity Searcher (HARPS) spectrograph under program 108.22LR.001 (PI: Psaridi) designed to specifically confirm and characterize exoplanets transiting AF-type stars. HARPS is mounted at the ESO 3.6m telescope in La Silla Observatory (\citealt{Mayor2003}). For several of our observations we used the HARPS high efficiency EGGS mode (R $\sim$ 80, 000) which is more suitable for hot stars with high projected rotational velocities since it allows a gain of two on the throughput and provides higher RV precision for photon limited observations. By using this mode one operates at lower resolution and higher instrumental throughput when compared with standard HARPS High Accuracy Mode (HAM). We obtained 23 observations of TOI-615 from 2021 November 10 to 2022 March 23. The adopted exposure time was 1200~s which translated into a S/N between 42 and 67. A total of 16 RVs were obtained for TOI-622 from 2021 October 12 to 2022 January 04 using a median exposure time of 1800~s, resulting in a S/N range of 106 to 183. For TOI-2641 we collected six RVs from 2022 March 12 to 2022 March 22 with an exposure time of 600~s and 1800~s and S/N between 41 and 52. Additionally, nine more spectra were collected for TOI-2641 as part of the NOMADS program (PI: Armstrong, 108.21YY.001) in High-Accuracy Mode (HAM, R $\sim$ 115, 000) from 2022 May 17 to 2022 May 27 with a median exposure time of 1800~s and S/N between 18 and 38. Since the HARPS observations of TOI-2641 were made under different modes they were treated as coming from independent instruments in our global analysis. The spectra were reduced with the standard calibration and reduction pipeline. The RVs were computed by cross-correlating with different binary masks and the one that reached higher precision in the mass measurement was selected. For TOI-615, our hottest star, we used an A0 binary mask while for TOI-622 and TOI-2641 we used a G2-type binary mask. For each star, we also computed the RVs with the Fourier interspectrum method (\citealt{Chelli:2000}), consisting of correlating, in the Fourier space, the spectra of the star with a reference spectrum built by averaging all the spectra acquired for this star (\citealt{Galland:2005}). The resulting RVs are consistent with those obtained with the CCF method described above. For TOI-615 one RV point (see crossed point in Figure \ref{fig:rvs}, left) was taken during transit (BJD 2459585.7779) and was affected by the Rossiter-McLaughlin (RM) effect with an expected semi-amplitude of 94 $\pm$ 12~\ms, therefore we excluded it from the analysis.

We monitored TOI-2641 with the high resolution CORALIE spectrograph that is installed at the Swiss 1.2-m Leonhard Euler Telescope at ESO’s La Silla Observatory \citep{Queloz2001}. CORALIE has a resolving power of R $\sim$ 60, 000 and is fed by a 2$\arcsec$ fiber (\citealt{Segrasan2010}). A total of 18 RVs were obtained from 2022 January 22 to 2022 March 15 using exposure times of 2400~s which translated in spectra with a signal-to-noise ratio per resolution element (S/N) between 17 and 25. We derived the RV of each epoch by cross-correlating the spectrum with a binary mask that matches the spectral type of the target \citep{Baranne1996}. The RVs of TOI-2641 were obtained using a G2 mask \citep{Pepe2002}. The bisector-span (BIS), the contrast (depth) and the full width at half-maximum (FWHM) were computed using the standard CORALIE data reduction pipeline. For each star, we also computed the RVs with the Fourier interspectrum method (see HARPS section above). These RVs are consistent with those obtained with the CCF method.

We obtained 32 spectra of TOI-2641 with the CTIO High Resolution Spectrometer (CHIRON; \citealt{Tokovinin2013}), a fiber-fed echelle spectrograph on the 1.5m telescope on Cerro Tololo. The spectra were collected between 2021 May 28 and 2022 December 13. The exposure time was between 1500 s or 1800 s, leading to a median S/N per extracted pixel of 20.
We observed using the image slicer mode, achieving a spectral resolution of $R\sim80000$, and we obtained a ThAr lamp immediately after the science exposure.
The spectra were reduced using the standard CHIRON pipeline \citep{Paredes2021}, and the radial velocities were computed following the method described in \citet{HD2685}.

\subsection{High resolution imaging}\label{sec:High resolution imaging}
The large TESS pixel size (21$\arcsec$) can result in possible flux contamination by nearby stars. This can cause flux dilution and therefore underestimation of the observed transit depth or even false-positive transit signals. Stellar companions can be ruled out with high angular resolution imaging.

\begin{figure*}
  \centering
  \includegraphics[width=0.32\textwidth]{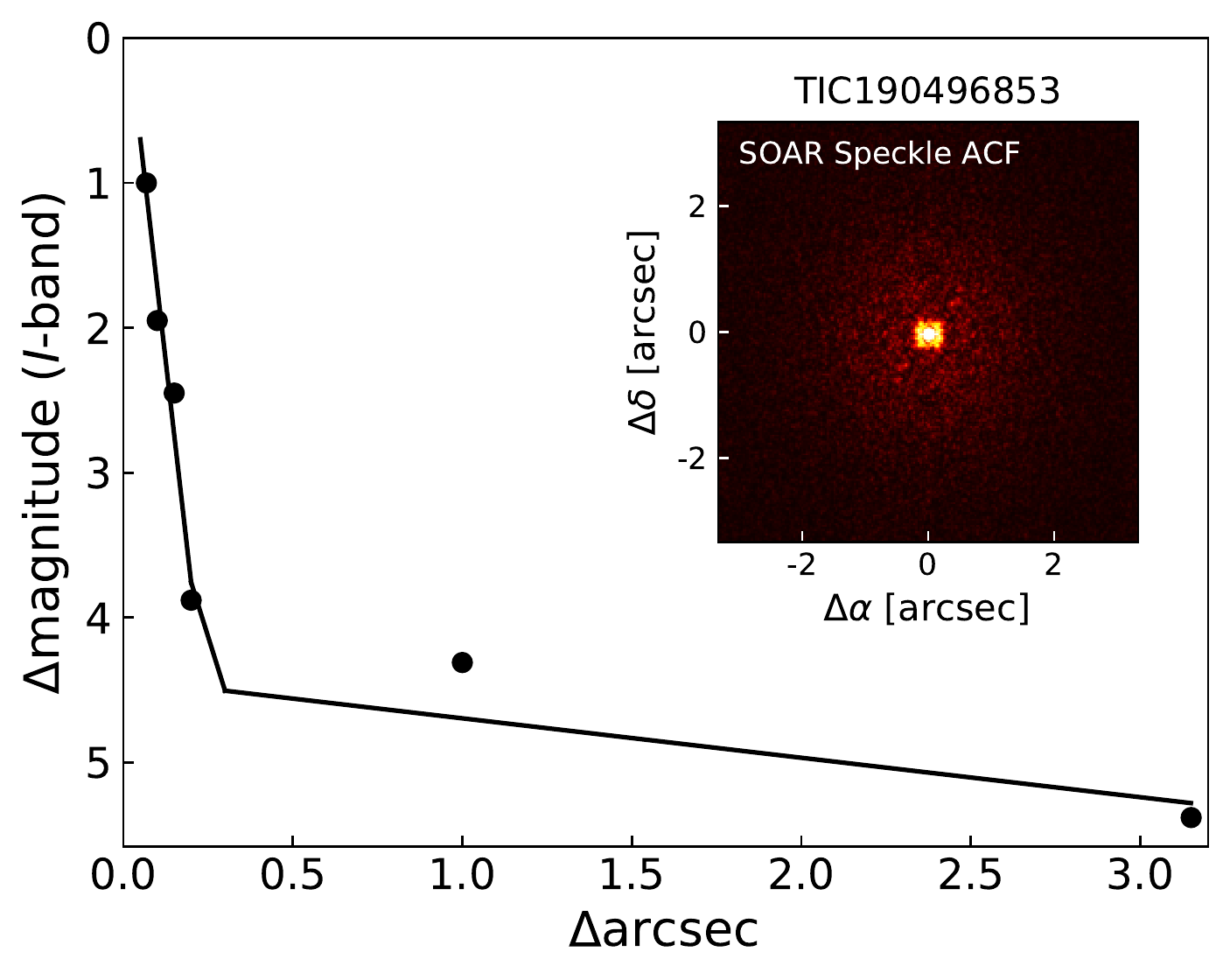}
  \hspace{0.2cm}
  \includegraphics[width=0.32\textwidth]{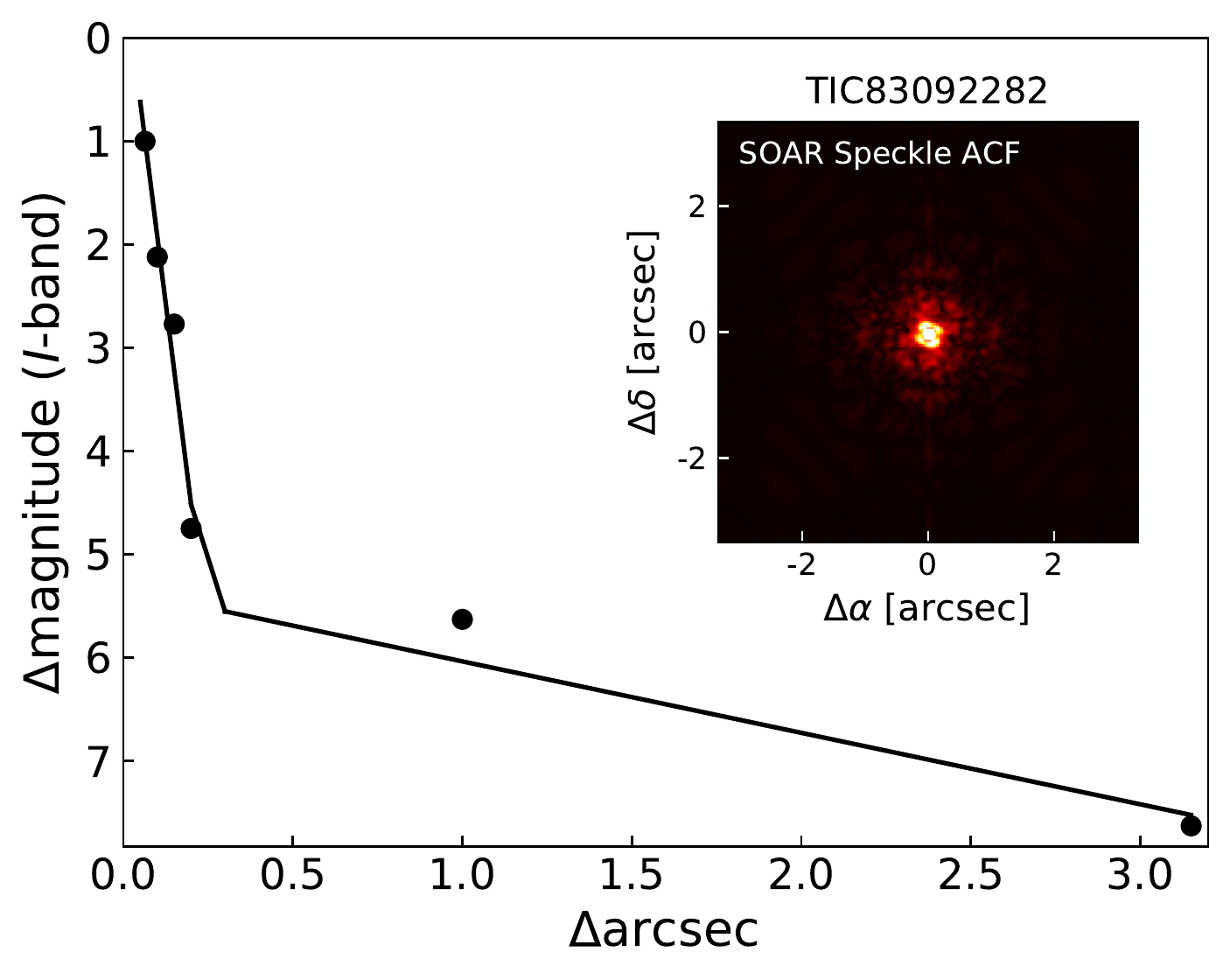}
  \includegraphics[width=0.313\textwidth]{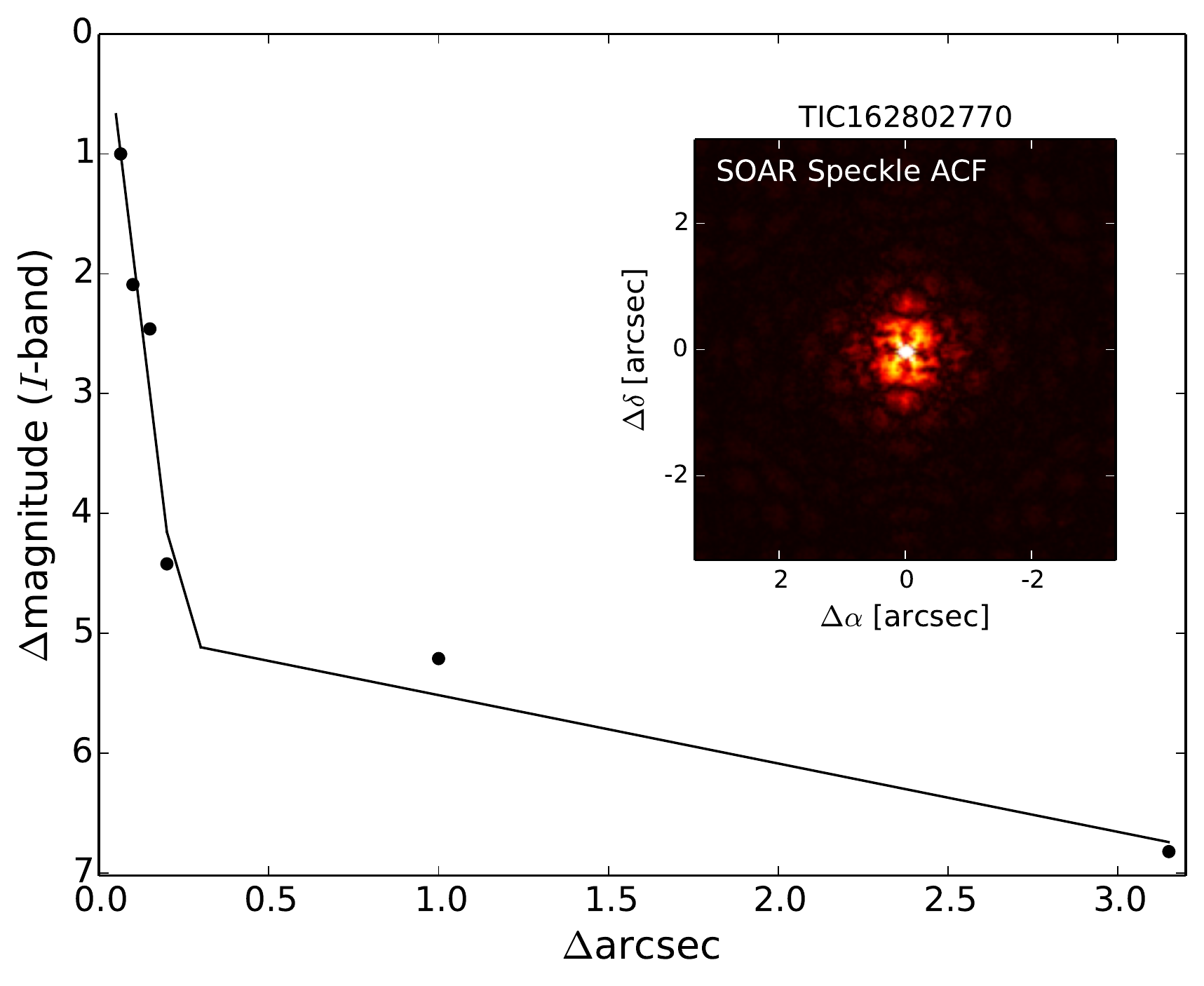}
  \caption{SOAR speckle imaging (5$\sigma$ upper limits) of TOI-615 (TIC 190496853; \textit{left}), TOI-622 (TIC 83092282; \textit{middle}), and TOI-2641 (TIC 162802770; \textit{right}) that rules out stellar companions within 3 $\arcsec$. The image inset represents the speckle auto-correlation function.} 
  \label{fig:SOAR}
\end{figure*}

We confirmed that there are no blended companions with speckle imaging on the 4.1-m Southern Astrophysical Research (SOAR) telescope \citep{SOAR}. The observations were taken in I-band, a similar visible bandpass as TESS. TOI-615 and TOI-622 were observed on 2019 May 18 and TOI-2641 on 2022 March 20 with no detection of nearby sources within 3$\arcsec$. The contrast curves with the 5$\sigma$ detection limit marked with a black line are shown in Figure~\ref{fig:SOAR}. The inset images zoomed and centered to the targets that represent the speckle auto-correlation function do not show any stellar companions.

\section{Analysis}\label{sec:analysis}
\subsection{Host star analysis}\label{sec:stellaranalysis}
\input{tables/stellarparameters}

\subsubsection{Spectral analysis}\label{sec:Spectral analysis}
For the spectral analysis of TOI-615 and TOI-622 we used the coadded HARPS spectra collected in EGGS mode (PI: Psaridi, 108.22LR.001). For the analysis of TOI-2641, we used the HARPS spectra collected in the high accuracy HAM mode (PI: Armstrong, 108.21YY.001). The spectral analysis was performed using methods similar to those given in \citet{2013MNRAS.428.3164D}. The effective temperature (\teff) was obtained from the excitation balance of Fe~{\sc i} lines and checked against that obtained from the H$\alpha$ line and from fitting the spectral energy distribution. Surface gravity was determined from the ionization balance between Fe~{\sc i} and Fe~{\sc ii} lines, with the Na~D lines being used as a check. The iron abundance relative to Solar, [Fe/H], was obtained using the equivalent widths of the Fe lines, with a value for microturbulence taken from the calibration of \cite{2012MNRAS.423..122B}. The Solar abundance of iron, log A(Fe) = 7.50,  was taken from \citet{2009ARA&A..47..481A}.

The projected stellar rotation velocity ($v \sin i_{*}$) was determined by fitting the profiles of several unblended Fe~{\sc i} lines in the wavelength region 600--620~nm.  For TOI-622 and TOI-2641 macroturbulence values of 5.9~km\,s$^{-1}$ and 4.7~km\,s$^{-1}$, respectively, were obtained from the asteroseismic calibration of \cite{2014MNRAS.444.3592D}. However, TOI-615 is too hot for this calibration, so a value of 7.4~km\,s$^{-1}$ was obtained by extrapolating the values given in Table~B.1 of \cite{2008oasp.book.....G}.

\subsubsection{Spectral energy distribution}\label{sec:SED}
\begin{figure*}
  \centering
  \includegraphics[width=0.36\textwidth]{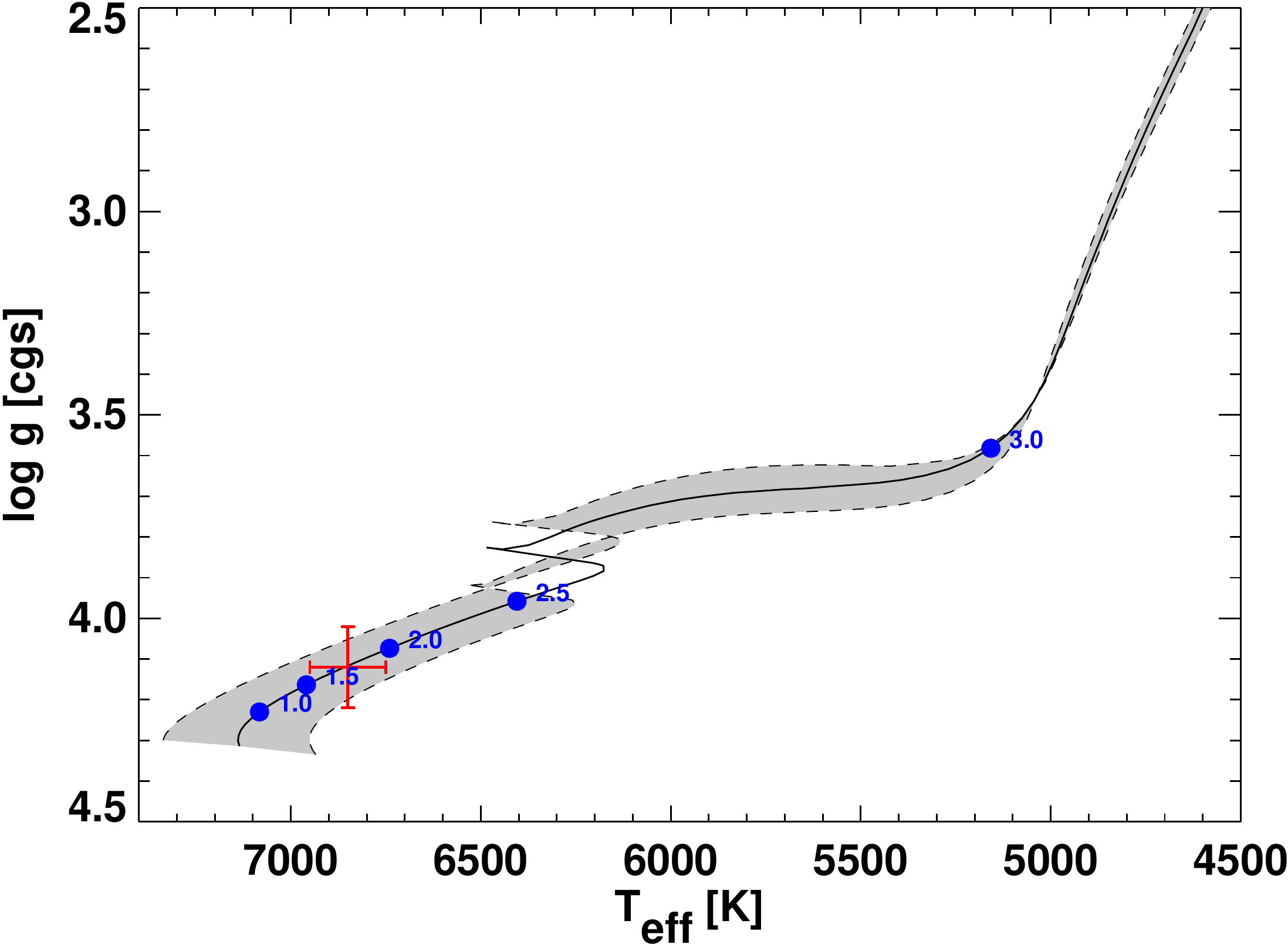}
  \hspace{1cm}
  \vspace{0.15cm}
  \includegraphics[width=0.36\textwidth]{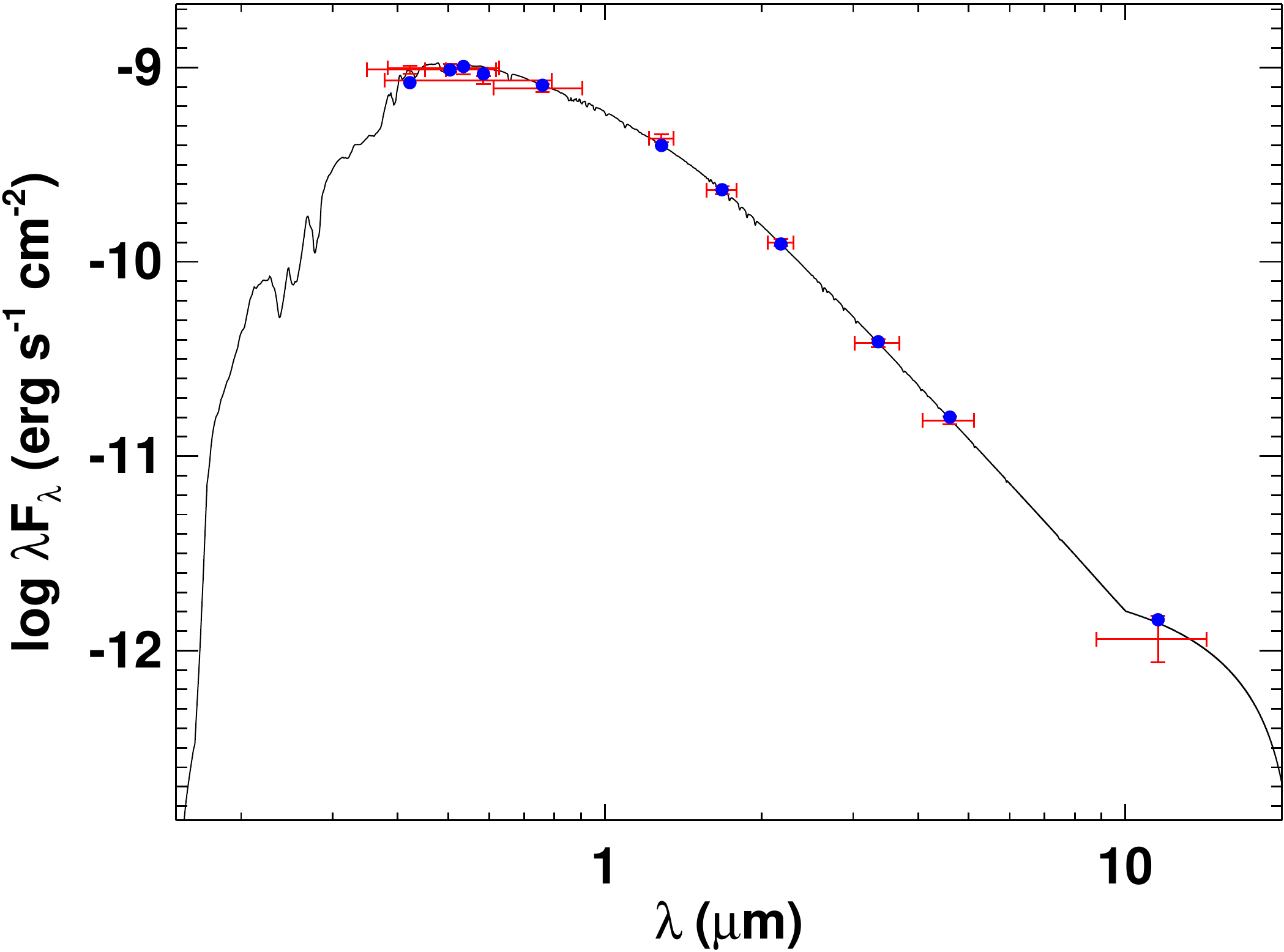} \\
  \vspace{0.15cm}
  \includegraphics[width=0.36\textwidth]{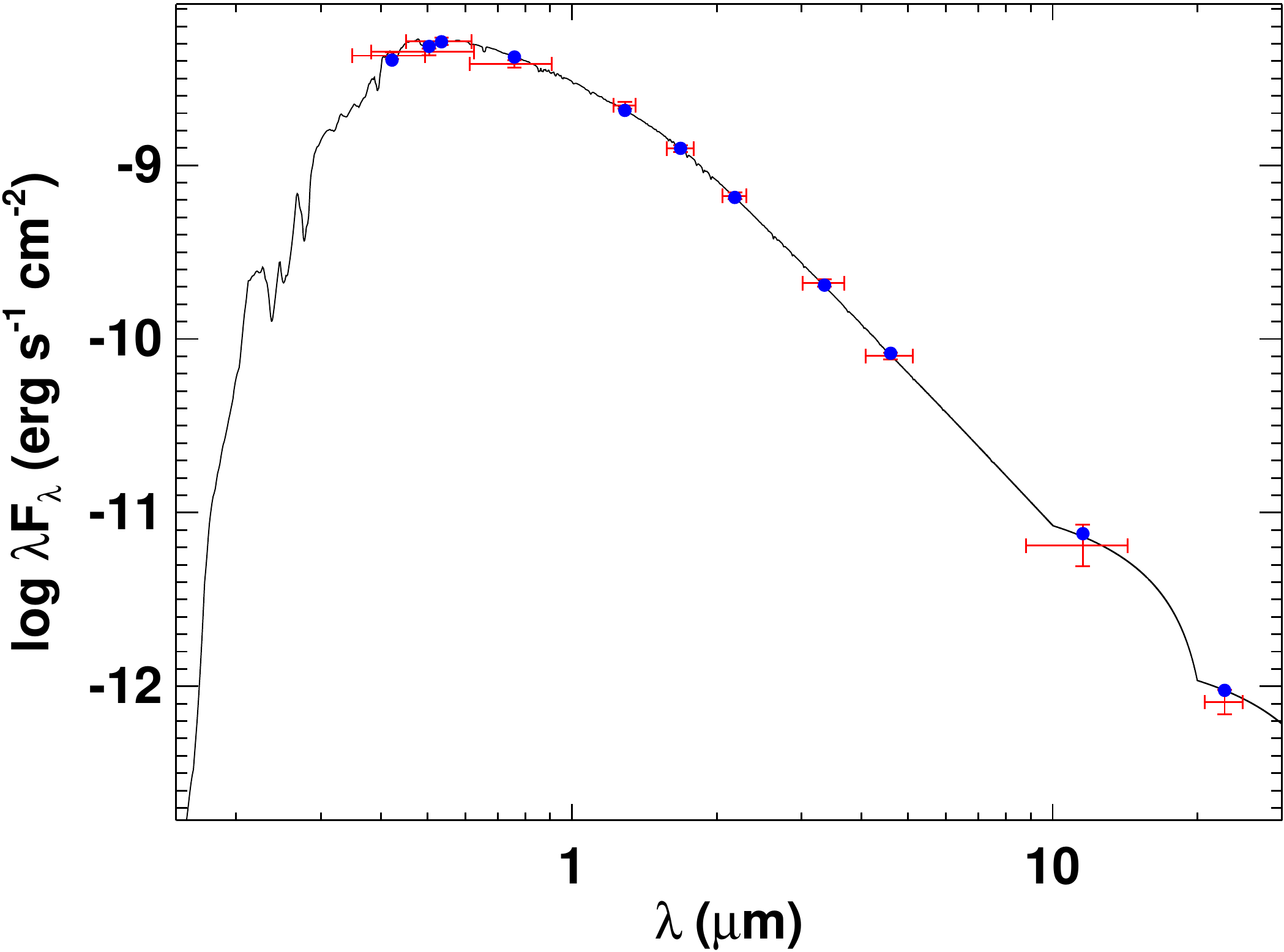}
  \hspace{1cm}
  \includegraphics[width=0.36\textwidth]{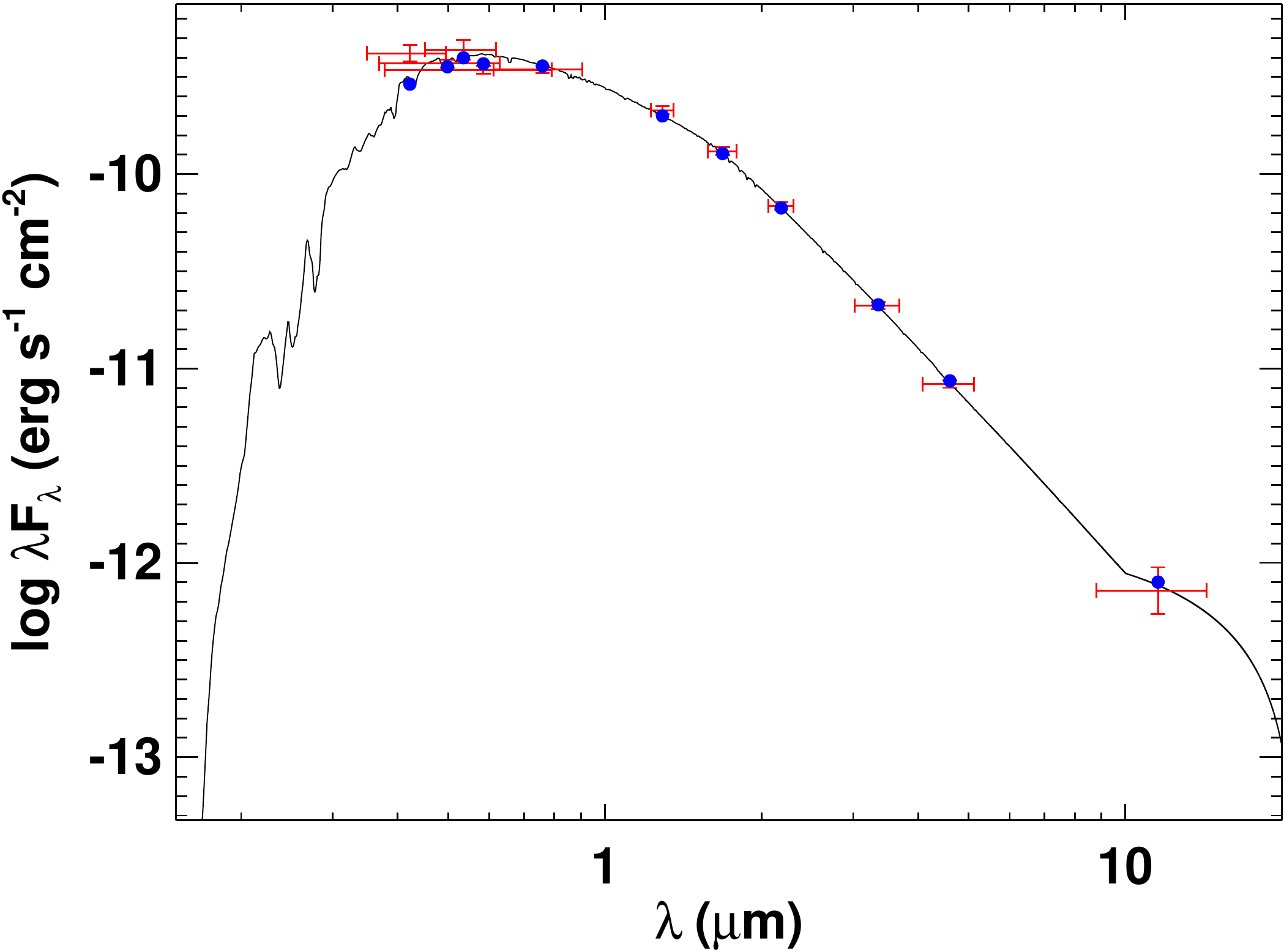}
  \caption{Kiel diagram of TOI-615 (\textit{upper left}) compared to model evolutionary track from \citet{Demarque:2004}. The red symbol represents the measured \teff~ and $\log g$, and the shaded swathe corresponds to the uncertainties in the stellar mass and metallicity. Age points are marked in blue. Spectral energy distribution of TOI-615 (\textit{upper right}), TOI-622 (\textit{bottom left}), and TOI-2641 (\textit{bottom right}). Red symbols represent the observed photometric measurements, where the horizontal bars represent the effective width of the passband. Blue symbols are the model fluxes from the best-fit Kurucz atmosphere model (black).} 
  \label{fig:sed}
\end{figure*}

As a photometric, empirical check for consistency with the spectroscopic and model based solution of the basic stellar parameters, we performed an analysis of the broadband spectral energy distribution (SED) of each star together with the {\it Gaia\/} EDR3 parallax \citep[with no systematic offset applied; see, e.g.,][]{StassunTorres:2021}, in order to determine an empirical measurement of the stellar radius, following the procedures described in \citet{Stassun:2016,Stassun:2017,Stassun:2018}. Depending on the photometry available for each source, we pulled the $B_T V_T$ magnitudes from {\it Tycho-2}, the $BVgri$ magnitudes from {\it APASS}, the $JHK_S$ magnitudes from {\it 2MASS}, the W1--W4 magnitudes from {\it WISE}, the $G G_{\rm BP} G_{\rm RP}$ magnitudes from {\it Gaia}, and the FUV and/or NUV fluxes from {\it GALEX}. Together, the available photometry generally spans the stellar SED over the approximate wavelength range 0.2--22~$\mu$m (see Figure~\ref{fig:sed}).  
We performed fits to the photometry using Kurucz stellar atmosphere models, with the principal parameters being the effective temperature (\teff), metallicity ([Fe/H]), and surface gravity ($\log g$), for which we adopted the spectroscopically determined values. We included the extinction, $A_V$, as a free parameter but limited to the full line-of-sight value from the Galactic dust maps of \citet{Schlegel:1998}. The resulting fits shown in Figure~\ref{fig:sed} have a reduced $\chi^2$ ranging from 1.1 to 1.6, and the best-fit parameters are summarized in Table~\ref{tab:Stellar parameters}. Integrating the model SED gives the bolometric flux at Earth, $F_{\rm bol}$. Taking the $F_{\rm bol}$ together with the {\it Gaia\/} parallax directly gives the luminosity, $L_{\rm bol}$. Similarly, the $F_{\rm bol}$ together with the \teff~ and the parallax gives the stellar radius, \rstar. Moreover, the stellar mass, \mstar, can be estimated from the empirical eclipsing-binary based relations of \citet{Torres:2010}, and the (projected) rotation period can be calculated from $R_\star$ together with the spectroscopically measured $v\sin i$. When available, the {\it GALEX} photometry allows the activity index, $\log R'_{\rm HK}$ to be estimated from the empirical relations of \citet{Findeisen:2011}. All quantities are summarized in Table~\ref{tab:Stellar parameters}.

Where possible, we have also estimated the system ages from the $R'_{\rm HK}$ activity and/or the stellar rotation, using the activity-age and/or rotation-age empirical relations of \citet{Mamajek:2008} which are applicable for \teff~$\lesssim$ 6500~K. These quantities are also summarized in Table~\ref{tab:Stellar parameters}. 
Note that in the case of TOI-622, it is included in the {\it Gaia}-based kinematic stellar age catalog of \citet{Kounkel:2020}, who estimate an age of $\sim 750$~Myr, consistent with the gyrochronological age estimate. Finally, in the specific case of TOI-615 for which neither a gyrochronological age nor an activity age estimate are possible, we compare the H-R diagram position of the star to the model isochrones of \citet{Demarque:2004}, from which we obtain an estimated age of  1.7 $\pm$ 0.3~Gyr (see Figure~\ref{fig:sed}, upper left).

\subsection{Joint transit and RV analysis}\label{sec:Global analysis}
We retrieved the planet parameters using a joint photometric and RV modeling with the \textbf{CO}de for transiting exopla\textbf{N}et
\textbf{AN}alysis (\CONAN, described in \citealt{Lendl2017}, \citeyear{toi222}), a Markov Chain Monte Carlo (MCMC) framework. \CONAN allows to jointly fit transit and RV data from different instruments while accounting for variations in the individual datasets using polynomial baseline models constructed from various external observational variables or Gaussian Processes (GPs). \CONAN incorporates GPs implemented using \george \citep[][]{george} and \celerite (\citealt{Foreman:2017}) packages, to tackle nonwhite noise and stochastic processes in the data. The fitted parameters in the analysis are as follows: orbital period, mid-transit time, radius ratio ($R_p$/$R_*$), impact parameter, transit duration, eccentricity, argument of periastron and RV amplitude. For limb darkening, we derived quadratic coefficients and their uncertainties for different photometric filters using the \ldcu$\footnote{\url{https://github.com/delinea/LDCU}}$ routine (\citealt{Deline2022}). We set a Gaussian prior on the TESS limb-darkening parameters and fixed them for the ground-based photometry due to the lower precision. To account for the instrumental, atmospheric and stellar correlated noise in the ground-based light curves, we used polynomials consisting of a combination of different variables (time, stellar FWHM, airmass, coordinate shifts, and sky background). The best combination of variables favored by the Bayesian Information Criterion is selected for each dataset (Table \ref{tab:tessandgroundphotometry}). For the analysis of the TESS data, we made use of the PDCSAP fluxes that are corrected for photometric dilution from neighbor stars, therefore our model does not include a dilution factor. To account for photometric variability and additional systematics, we detrended the light curves with GP using the approximate Mat\'ern-3/2 kernel (\citealt{Foreman:2017}) included in \celerite package. 

For the RV fitting, we ran the MCMC analysis both with the eccentricity as a free parameter
and fixed to zero and found that the results are compatible with a circular orbit. Additionally, we estimated the significance of the nonzero eccentricity in order to avoid Lucy-Sweeney bias (\citealt{Lucy1971}). Using the $F$-test approach (\citealt{Lucy1971}) we find a 60$\%$, 57$\%$, and 5$\%$ probability for TOI-615, TOI-622, and TOI-2641, respectively, that the improvement in the fit could have arisen by chance if the orbit were circular. Consequently we fixed the eccentricity to zero in the global analysis. The priors used in the MCMC analysis and the posterior estimates can be found in Table~\ref{tab:CONAN priors} and Table~\ref{tab:CONAN results}, respectively.

\begin{figure}[ht]
  \centering
  \includegraphics[width=0.45\textwidth]{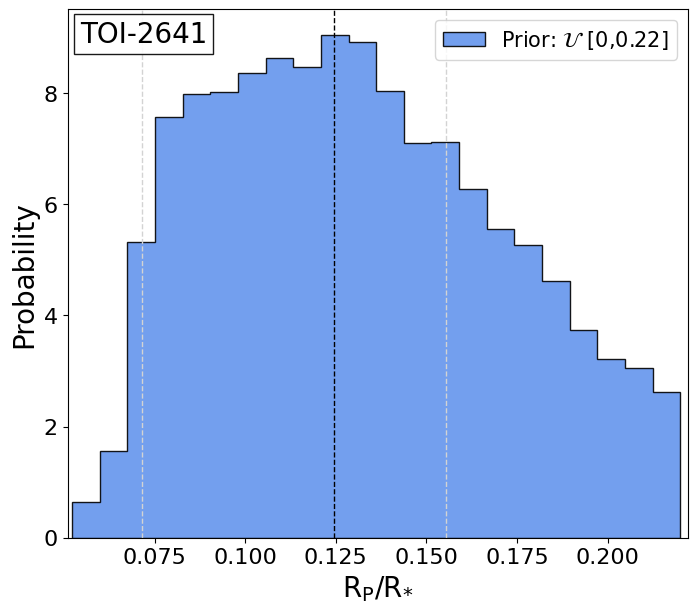}
  \caption{\CONAN posterior distribution of the R$_{P}$/R$_{*}$ of TOI-2641. The black vertical line corresponds to the maximum value and the two gray vertical lines indicate the 68$\%$ credible intervals.} 
  \label{fig:grazinghistogram}
\end{figure}

As mentioned in Section \ref{sec:TESS_photometry}, the V-shaped light curves of TOI-2641b suggest a grazing geometry. For the analysis, we set a broad uniform prior on the impact parameter of $\mathcal{U}(0, 1.5)$ and a large prior on $R_{\mathrm{P}}/R_{\mathrm{*}}$ of $\mathcal{U}(0, 0.22)$ assuming a maximum planetary radius of $R_{\mathrm{P}}$ $<$ 3 \rjup. The strong degeneracy between the impact parameter $b$ and the companion radius $R_{\mathrm{P}}$ results in skewed posterior distributions (Figure \ref{fig:grazinghistogram}). Given this asymmetry, we extracted the best fitting values and their uncertainties using the mode (maximum value) and FWHM of each distribution. The grazing configuration results in large uncertainties on the planetary radius.

Finally, we performed a frequency analysis on the TESS light curves in order to identify stellar oscillations and retrieve the stellar rotation periods. For the analysis we made use of the TESS SAP-flux light curves and calculated the Lomb-Scargle periodograms (\citealt{Lomb}, \citealt{Scargle}) after masking the transit events. None of the respective periodograms revealed significant signals around the rotation periods identified by the spectral analysis.

\input{tables/planetparamconan}
\subsection{Planetary evolution models}
Using CEPAM (\citealt{guillot1995}) and a nongray atmosphere (\citealt{parmentier2015}), we model the evolution of TOI-615b, TOI-622b, and TOI-2641b. We assume simple structures consisting of a central dense core surrounded by a hydrogen and helium envelope of solar composition. The core is assumed to be composed of 50$\%$ ices and 50$\%$ rocks, by mass. Since all these planets are highly irradiated, we account for the dissipation of energy in the interior, following the approach of \citealt{guillot2002}: A fraction $\epsilon$ of the irradiation luminosity $L_\star$ is assumed to be dissipated at the bottom of the envelope and included in the evolution calculations. 

\begin{figure}[ht]
  \centering
  \includegraphics[width=0.5\textwidth]{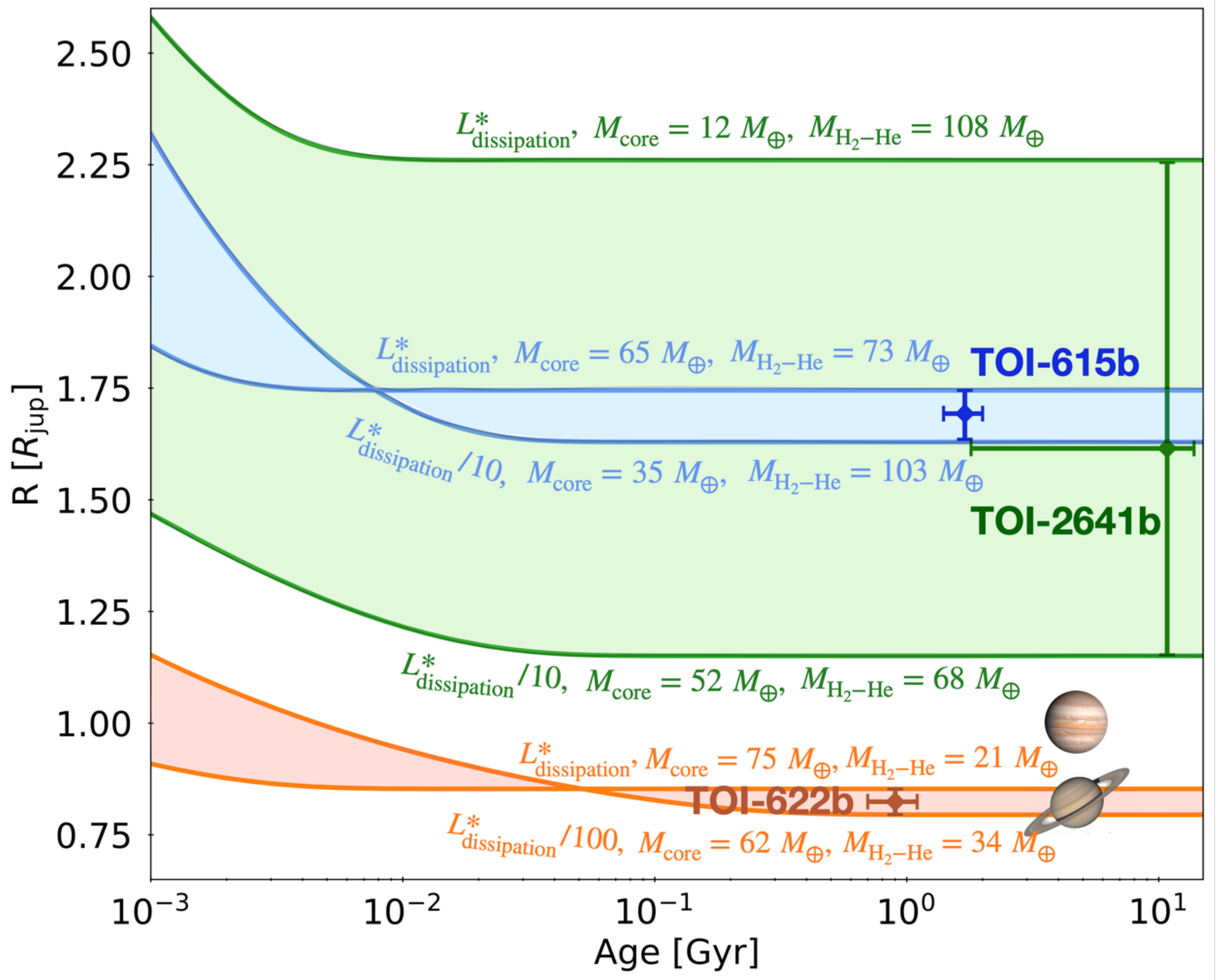}
  \caption{Evolution models of TOI-615b, TOI-622b, and TOI-2641b. All models assume a central ice-rock core surrounded by a hydrogen-helium envelope of solar composition. Models account for dissipation of energy in the interior due to important irradiation and $L_{\rm dissipation}^*$ corresponds to our fiducial dissipation luminosity. For each planet, the ranges of core masses and envelope masses are shown. The errorbars correspond to observational constraints on their age and radius, compared to the ones of Jupiter and Saturn.} 
  \label{fig:evolution_models}
\end{figure}

The source and magnitude of this dissipation is still under investigation, but using a statistical approach, \citealt{thorngren2018} estimate for exoplanets with these irradiation levels that $\epsilon\sim 2\%$. This yields a luminosity due to internal dissipation $L_{\rm dissipation}^*=10^{28}$, $10^{27}$ and $5 \times 10^{27}\,$erg/s, for TOI-615b, TOI-622b, and TOI-2641b respectively. In Figure~\ref{fig:evolution_models}, we explore how observational constraints and uncertainties on dissipation affect what we can infer on the bulk composition of these planets. We show the results of calculations with our fiducial dissipation luminosity compared to models with $L_{\rm dissipation}^*$ divided by 10 (for TOI-615b and TOI-2641b) or 100 (for TOI-622b). After a few to hundred million years, our models predict that the planetary radii remain constant, due to internal heating that exceeds the planet's cooling luminosity.

We find that these planets tend to be significantly enriched in heavy elements. Defining an approximate bulk metallicity as $Z\approx M_{\rm core}/M_{\rm tot}$ and assuming $Z_\odot=0.015$, we find $Z/Z_\odot=17$ to 31 for TOI-615b, 43 to 52 for TOI-622b, and 7 to 29 for TOI-2641b. For comparison, Jupiter and Saturn have bulk metallicities $Z/Z_\odot$ ranging from 4 to 14, depending on assumptions on interior models (\citealt{guillot2022}). For TOI-615b, and to some extent for TOI-622b, heat dissipation is found to be the dominant factor controlling the inferred composition. Spectroscopic observations of these planets (e.g., with JWST) would be highly valuable for the possibility to constrain the atmospheric metallicity and relate it to the bulk metallicity.

We also modeled the three targets using the planetary evolution code \texttt{completo21} (\citealt{Mordasini2012}). We used the implementation of \texttt{completo21} done by \citet{Sarkis2021}. \citet{Sarkis2021} used a Bayesian framework to couple the interior structure to the observed properties of hot-Jupiters. The planets are modeled with no central core and the heavy elements are assumed to be homogeneously mixed in the interior. The heavy elements are modeled as water with the equation of state ANEOS (\citealt{Thompson1990}; \citealt{Mordasini2020}) and hydrogen and helium are modeled with the SCvH equation of state (\citealt{Saumon1995}). The code uses a grid of fully nongray atmospheric models from \texttt{petitCODE} (\citealt{Molliere2015}, \citeyear{Molliere2017}) and the species considered are listed in \citet{Sarkis2021}. We also assume that part of the stellar irradiation is transported into the planet interior. The statistical model infers the internal luminosity of the planet and thus the heating efficiency. We find an internal luminosity equal to $9\times10^{27}$, $7 \times 10^{26}\,$, and $5.5 \times 10^{27}\,$erg/s for TOI-615b, TOI-622b, and TOI-2641b respectively. Assuming the same stellar metallicity of $Z_\odot=0.015$, we can derive the planet metal enrichment $Z/Z_\odot$ and find values in agreement with the ones derived with CEPAM:  $19\pm11$ for TOI-615b, $43\pm5$ for TOI-622b, and $15\pm9$ for TOI-2641b. We show that considering a central core or not does not change significantly the results in terms of planet metal enrichment and both models are able to well reproduce the observed radii.

\clearpage
\newpage

\section{Discussion and conclusion}\label{sec:Discussion and conclusion}
\begin{figure*}[ht!]
  \centering
  \includegraphics[width=0.97\textwidth]{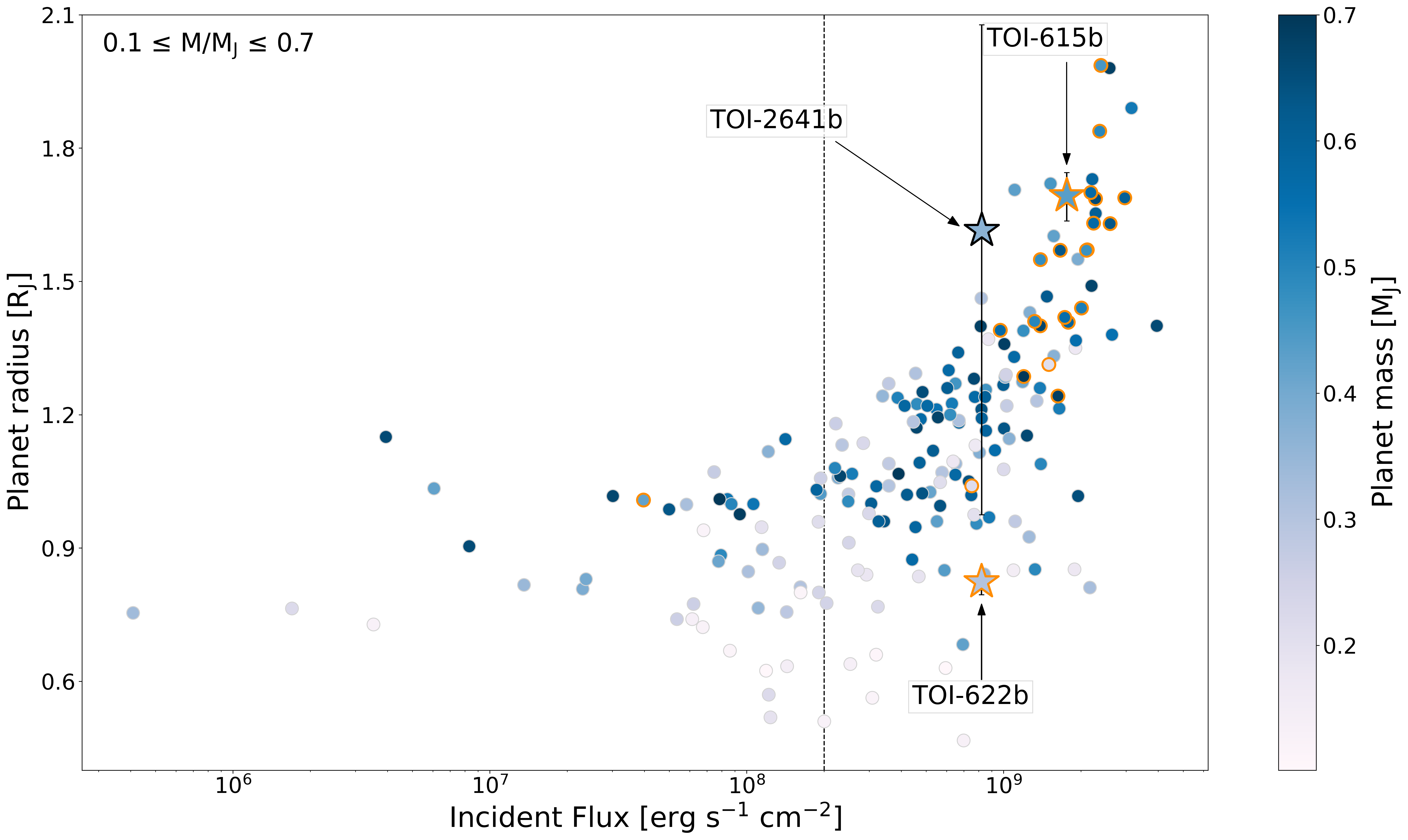}
  \caption{Planet radius over incident flux for known giant (R $>$ 5~\re) planets within the mass interval 0.1-0.7 \mjup. The points are colored by the planetary mass. The vertical dashed line corresponds to the planetary inflation threshold (146 F$_{\oplus}$, \citealt{demoryandseager}). Stars with effective temperatures above the Kraft break are outlined in orange. The three new planets, TOI-615b, TOI-622b, and TOI-2641b, have been highlighted as stars and labeled.} 
  \label{fig:radiusincident}
\end{figure*}

We have begun a RV survey to search for planets and brown dwarfs around AF-type stars using TESS data (\citealt{Psaridi2022}). As part of this survey, we present the detection and characterization of three new Saturn-mass planets that are transiting F-type stars, TOI-615b, TOI-622b, and TOI-2641b. Our analysis is based on TESS 2-minute cadence and FFI data from the first three years of its mission, ground-based follow-up photometry from EulerCam, NGTS, LCOGT, ASTEP, and El Sauce. For the analysis, we fit the photometric data jointly with RV data from HARPS, CORALIE, and CHIRON using \CONAN. 

TOI-615 and TOI-622 have effective temperatures above the Kraft break with values of \teff = 6850 $\pm$ 100~K and \teff = 6400 $\pm$ 100~K, respectively. Previous to our survey, only 22 Saturn-mass exoplanets (0.1-0.7 \mjup) have been confirmed to be transiting stars with \teff > 6200 K. TESS has contributed in this sample with four new detections (TOI-201b; \citealt{toi201}, TOI-1842b; \citealt{TOI1842}), including TOI-615b and TOI-622b. 
In Figure \ref{fig:radiusincident} we illustrate the planetary radius over the incident flux for all known, well-characterized transiting giant (R $>$ 5 \re) planets within the mass interval 0.1-0.7 \mjup. The points are colored by the planetary mass and we highlight the planets around stars above the Kraft break in orange and the planets confirmed in this work as stars. The vertical line corresponds to the observed threshold for inflation (\citealt{demoryandseager}) which is based on the $\textit{Kepler}$ sample, showing that transiting hot Jupiters that receive a level of incident flux larger than $\sim$2~$\times$ 10$^8$~erg~s$^{-1}$~cm$^{-2}$ ($\sim$146 $F_{\oplus}$) appear to experience radius inflation. As seen in the diagram, planets with masses in the range of $\sim$0.1-0.4 \mjup, generally have radii R < 1.5 \rjup while more massive planets are found to have radii as large as $\sim$ 1.9 \rjup. This correlation between high irradiation, large mass and inflated radii has been established by \citealt{Sest2018}. The constrained masses and radii of TOI-615b, TOI-622b, and TOI-2641b follow this correlation.

 With a mass of 0.44 $\pm$ 0.08~\mjup and a radius of 1.69 $\pm$ 0.05~\rjup, TOI-615b has a density of 0.119 $\pm$ 0.026 g/cm$^3$ ($\sim$0.09~\densjup), thus it is ranked among the 2$\%$ lowest density, well-characterized known planets and it the least-dense planet discovered with TESS. Moreover, TOI-615b is the lowest density planet that transits a star with such high effective temperature (\teff = 6850 $\pm$ 100 K). Placing TOI-615b on the radius-incident flux diagram (Figure \ref{fig:radiusincident}) we see that the planet is among the largest hot Saturns known. We calculate the incident flux received by TOI-615b to be $\sim$1.75~$\times$ 10$^9$~erg~s$^{-1}$~cm$^{-2}$ ($\sim$1277 $F_{\oplus}$) which is one order of magnitude larger than the insolation threshold for inflation.  
Using the \citet{Weiss2013} empirical relationship for predicted radius of planets with M$_P$ $<$ 150 \me (M$_P$ $<$ 0.47~\mjup) that makes use of the planetary mass and incident flux:
$$
\frac{R_{\mathrm{P}}}{R_{\oplus}}=1.78\left(\frac{M_{\mathrm{P}}}{M_{\oplus}}\right)^{0.53}\left(\frac{F}{\operatorname{erg~s}^{-1} \mathrm{~cm}^{-2}}\right)^{-0.03}
$$
we predict a planetary radius  of 12.8 $\pm$ 1.3~\re. With a measured radius of 18.6 $\pm$ 0.6~\re, TOI-615b appears to be inflated by $\sim$~45$\%$. 

In contrast to TOI-615b, TOI-622b with a mass of 0.29 $\pm$ 0.07~\mjup and a radius of 0.83 $\pm$ 0.03~\rjup is unusually dense with a density of 0.718 $\pm$ 0.181 g/cm$^3$ ($\sim$0.52~\densjup). TOI-622b receives an incident flux of $\sim$8.2~$\times$ 10$^8$~erg~s$^{-1}$~cm$^{-2}$ ($\sim$600 $F_{\oplus}$) which is well above the lower inflation limit, however, the planet does not show evidence of inflation. TOI-622b has a predicted planetary radius of 10.8 $\pm$ 1.3~\re; its observed radius of 9.04 $\pm$ 0.31~\re is thus slightly smaller (1.3$\sigma$) than expected. 

TOI-2641b is a 0.39 $\pm$ 0.04~\mjup hot Jupiter with a poorly constrained radius (1.61$^{+0.46}_{-0.64}$ \rjup) due to the extremely grazing nature of the transit with impact parameter of $b$ = 1.042$^{+0.048}_{-0.065}$. The V-shaped transits obtained by TESS, EulerCam, and LCOGT can be seen in Figure \ref{fig:Light curves}. As shown in Figure \ref{fig:radiusincident}, TOI-2641b receives 613 times more incident flux than Earth. Given its mass and insolation, we do not expect TOI-2641b to have a radius larger than $\sim$ 1.5 \rjup. This planet adds to the population of grazing planets ($b$+$R_{p}$/$R_{*}$ $\geq$ 1) with 19 confirmed detections to-date. These targets have sensitive transit depth and duration which makes them promising for long term monitoring in searches for additional planets through TTVs and TDVs (\citealt{Ribas2008}).

As discussed in Section \ref{sec:Global analysis}, we modeled all three systems assuming a circular orbit. We estimated the circularization timescales using equation 3 of \citet{Laughlin06} for tidal quality factor $Q_p$ = 10$^\mathrm{5}$ and 10$^\mathrm{6}$. TOI-615b might have circularized after 0.021$^{+0.008}_{-0.007}$~Gyr for $Q_p$ = 10$^\mathrm{5}$ and 0.21$^{+0.08}_{-0.07}$~Gyr for $Q_p$ = 10$^\mathrm{6}$, which is shorter than the lifetime of the system (1.7 $\pm$ 0.3~Gyr). For TOI-622b the expected circularization timescale is 0.81$^{+0.46}_{-0.51}$~Gyr using $Q_p$ = 10$^\mathrm{5}$ and 8.1$^{+4.6}_{-5.1}$~Gyr using $Q_p$ = 10$^\mathrm{6}$. Compared to the age of the host star (0.9 $\pm$ 0.2~Gyr), TOI-622b might have undergone tidal
circularization or formed in situ in a nearly circular orbit. Finally, the expected tidal circularization timescale of TOI-2641b is 0.016$^{+0.024}_{-0.016}$~Gyr for $Q_p$ = 10$^\mathrm{5}$ and 0.16$^{+0.24}_{-0.16}$~Gyr for $Q_p$ = 10$^\mathrm{6}$ which is significantly shorter than the estimated stellar age (10.8 $\pm$ 9.0~Gyr). This is consistent with TOI-2641b having undergone circularization before reaching its present age.

\begin{figure}[ht]
  \centering
    \includegraphics[width=0.5\textwidth]{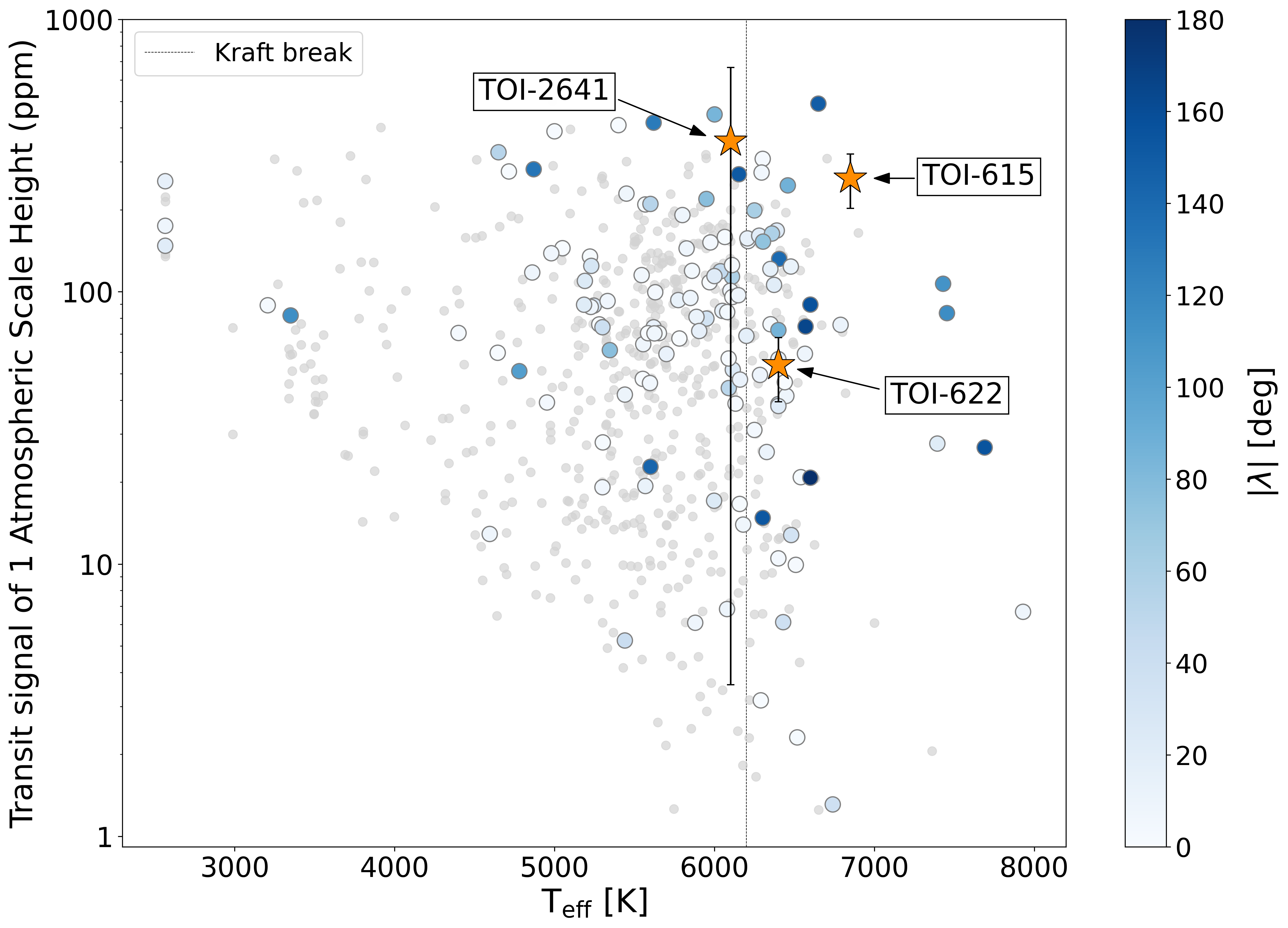}
  \caption{Expected transmission spectral signal of 1 atmospheric scale height (\citealt{Bento2014}) versus stellar effective temperature for exoplanets with precise densities ($\sigma_{M}/M$ $\leq$ 25$\%$ and $\sigma_{R}/R$ $\leq$8$\%$). The points of planets with Rossiter–McLaughlin (RM) observations are color-coded with the absolute value of their sky-projected spin-orbit angle ($|\lambda|$). The three new planets, TOI-615b, TOI-622b, and TOI-2641b, have been highlighted as orange stars and labeled.} 
  \label{fig:Scale height}
\end{figure}

\citet{winn2010} noted that planets orbiting stars with effective temperatures above the Kraft break have a higher probability to be in a polar or retrograde orbit. In fact, 20 out of the 35 planets on misaligned orbits ($|\mathrm{\lambda}|$>10$^{\circ}$, with >3$\sigma$ confidence) have been found to orbit hot stars (Figure \ref{fig:Scale height}). Studies of the projected spin-orbit angle of can bring essential insights on the formation processes and migration of planets transiting massive, early-type stars. Given the high effective temperature, the large planetary radius and high stellar rotational velocity, TOI-615 and TOI-622 are very appealing systems for future RM effect observations (\citealt{Rossiter1924}; \citealt{McLaughlin1924}). We predict the semi-amplitude of the RM effect for each planet by using Equation 1 in \citet{Triaud2018}. For TOI-615b, the expected semi-amplitude of the RM signal is A$_{RM}$ = 92 $\pm$ 10 \ms and for TOI-622 A$_{RM}$ = 142 $\pm$ 20 \ms. As discussed in Section \ref{sec:Confirmation spectroscopy}, one data point of TOI-615 occurred during transit and was affected by the RM effect with a semi-amplitude of 94 $\pm$ 12~\ms providing weak constrain on the orbital obliquity. 

The large predicted signals of TOI-615 and TOI-622 suggest that the targets are good candidates for such observations. As the RM technique relies on isolating the stellar signal emanating from the regions occulted by the planets, grazing transits are suboptimal for exploiting the RM effect. Plus, the majority of the techniques employed to analyze this effect (e.g., velocimetry, \citealt{Ohta2005}; Doppler tomography, \citealt{Gandolfi2012}) assume that the successive occulted stellar profiles are constant during the transit, which naturally biases the derived parameters (\citealt{Cegla2016}, \citealt{Bourrier2017}) even in fiducial cases. This assumption is all the less justifiable in case of a grazing transit and would accentuate those biases. Consequently, such studies could be challenging for TOI-2641.

Due to the high equilibrium temperature of the planets, we investigate the possibility of future atmospheric studies via transmission spectroscopy. Figure \ref{fig:Scale height} shows the expected transmission spectral signal of 1 atmospheric scale height ($H$, assuming a mean molecular weight of 2.2 g/mol) approximated by \citet{Bento2014}:
\begin{equation}
\centering
A_\mathrm{abs}=\frac{2 R_{\mathrm{p}} H}{R_{\star}^2}
\label{eq:Aabs}
\end{equation}
as a function of the stellar effective temperature for exoplanets with well-constrained densities. The expected absorption signals are: 261$^{+59}_{-58}$~ppm, 54$^{+13}_{-14}$~ppm, 358$^{+309}_{-354}$~ppm for TOI-615b, TOI-622b, and TOI-2641b respectively. Figure \ref{fig:Scale height} locates TOI-615b in the 2$\%$ of planets with highest A$_\mathrm{abs}$ that makes it an ideal candidate for future atmospheric studies. Additionally, observations with the James Webb Space Telescope (JWST) can allow the comparison of planets that suffer from extreme stellar irradiation to cooler gas giants. For that purpose, we calculate the Transmission Spectroscopy Metric (TSM, \citealt{Kempton2018}) that corresponds to the expected transmission spectroscopy signal-to-noise ratio (S/N). A higher TSM implies easier determination of the planet’s atmospheric spectrum.
We computed the TSM from \citealt{Kempton2018}:
\begin{equation}
\centering
\mathrm{TSM}=\textit {S} \times \frac{R_{P}^{3} T_{eq}}{M_{P} R_{*}^{2}} \times 10^{-m_{J} / 5},
 \label{eq:TSM}
\end{equation}
where $m_{J}$ is the apparent magnitude of the star in the J band. The scale factor ($S$) depends on the planetary radius as: $S=0.19$ for $R_{\mathrm{P}}<1.5~R_{\oplus};~ S=1.26$ for $1.5<R_{\mathrm{P}}<2.75~ R_{\oplus};~ S=1.28$ for $2.75<R_{\mathrm{P}}<4.0~R_{\oplus}$ and $S=1.15$ for $4.0<R_{\mathrm{P}}<10~R_{\oplus}$. For TOI-615b and TOI-2641b we adopted a scale factor $S$ = 1 since their radii are larger than 10~$R_{\oplus}$. For TOI-615b we calculate a value of TSM = 286 $\pm$ 19, for TOI-622b TSM = 203 $\pm$ 14, and for TOI-2641b TSM = 351$^{+303}_{-351}$. According to Table 1 from \citet{Kempton2018}, TOI-615b and TOI-622b are ranked in the first quartile, and hence they have beneficial properties for follow-up atmospheric characterization through transmission spectroscopy with $\textit{JWST}$ with TOI-615b being especially promising.

\begin{acknowledgements}
We thank the Swiss National Science Foundation (SNSF) and the Geneva University for their continuous support to our planet low-mass companion search programs. This work was carried out within the framework of the Swiss National Centre for Competence in Research (NCCR) $PlanetS$ supported by the Swiss National Science Foundation (SNSF) under grants 51NF40$\_$182901 and 51NF40$\_$205606. ML and BA acknowledge support of the Swiss National Science Foundation under grant number PCEFP2$\_$194576. This publication makes use of The Data $\&$ Analysis Center for Exoplanets (DACE), which is a facility based at the University of Geneva (CH) dedicated to extrasolar planet data visualization, exchange, and analysis. DACE is a platform of NCCR $PlanetS$ and is available at https:$//$dace.unige.ch. This paper includes data collected by the TESS mission. Funding for the TESS mission is provided by the NASA Explorer Program. This research has made use of the Exoplanet Follow-up Observation Program (ExoFOP; DOI: 10.26134/ExoFOP5) website, which is operated by the California Institute of Technology, under contract with the National Aeronautics and Space Administration under the Exoplanet Exploration Program. Funding for the TESS mission is provided by NASA's Science Mission Directorate. KAC acknowledges support from the TESS mission via subaward s3449 from MIT. We acknowledge the use of public TESS data from pipelines at the TESS Science Office and at the TESS Science Processing Operations Center. Resources supporting this work were provided by the NASA High-End Computing (HEC) Program through the NASA Advanced Super computing (NAS) Division at Ames Research Center for the production of the SPOC data products. The material is based upon work supported by NASA under award number 80GSFC21M0002. Based in part on observations obtained at Cerro Tololo Inter-American Observatory at NSF’s NOIRLab (NOIRLab Prop. ID 2022A-932354; PI: C. Ziegler), which is managed by the Association of Universities for Research in Astronomy (AURA) under a cooperative agreement with the National Science Foundation, and observations obtained at the Southern Astrophysical Research (SOAR) telescope, which is a joint project of the Ministério da Ciência, Tecnologia e Inovações (MCTI/LNA) do Brasil, the US National Science Foundation’s NOIRLab, the University of North Carolina at Chapel Hill (UNC), and Michigan State University (MSU). JSJ acknowledges support by FONDECYT grant 1201371 and partial support from the ANID Basal project FB210003. This research received funding from the European Research Council (ERC) under the European Union's Horizon 2020 research and innovation programme (grant agreement n$^\circ$ 803193/BEBOP), and from the Science and Technology Facilities Council (STFC; grant n$^\circ$ ST/S00193X/1). DJA is supported by UK Research \& Innovation (UKRI) through the STFC (ST/R00384X/1) and EPSRC (EP/X027562/1).
MINERVA-Australis is supported by Australian Research Council LIEF Grant LE160100001, Discovery Grants DP180100972 and DP220100365, Mount Cuba Astronomical Foundation, and institutional partners University of Southern Queensland, UNSW Sydney, MIT, Nanjing University, George Mason University, University of Louisville, University of California Riverside, University of Florida, and The University of Texas at Austin.
We respectfully acknowledge the traditional custodians of all lands throughout Australia, and recognize their continued cultural and spiritual connection to the land, waterways, cosmos, and community. We pay our deepest respects to all Elders, ancestors and descendants of the Giabal, Jarowair, and Kambuwal nations, upon whose lands the Minerva-Australis facility at Mt Kent is situated. The postdoctoral fellowship of KB is funded by F.R.S.-FNRS grant T.0109.20 and by the Francqui Foundation. This research has used data from the CTIO/SMARTS 1.5m telescope, which is operated as part of the SMARTS Consortium by RECONS (\url{www.recons.org}) members Todd Henry, Hodari James, Wei-Chun Jao, and Leonardo Paredes. At the telescope, observations were carried out by Roberto Aviles and Rodrigo Hinojosa. The CHIRON data were obtained from telescope time allocated under the NN-EXPLORE program with support from the National Aeronautics and Space Administration (Prop. IDs 2021B-0162, 2022A-543544, PI: Yee). This work makes use of observations from the LCOGT network. Part of the LCOGT telescope time was granted by NOIRLab through the Mid-Scale Innovations Program (MSIP). MSIP is funded by NSF. This work makes use of observations from the ASTEP telescope. ASTEP benefited from the support of the French and Italian polar agencies IPEV and PNRA in the framework of the Concordia station program, from OCA, INSU, Idex UCAJEDI (ANR- 15-IDEX-01), and ESA through the Science Faculty of the European Space Research and Technology Centre (ESTEC).

\end{acknowledgements}
\newpage
\bibliographystyle{aa}
\typeout{}
\bibliography{bib}
\begin{appendix}
\appendix
\onecolumn
\newpage
\section{Light curves}

\begin{figure}[ht]
  \centering
  \includegraphics[width=0.32\textwidth]{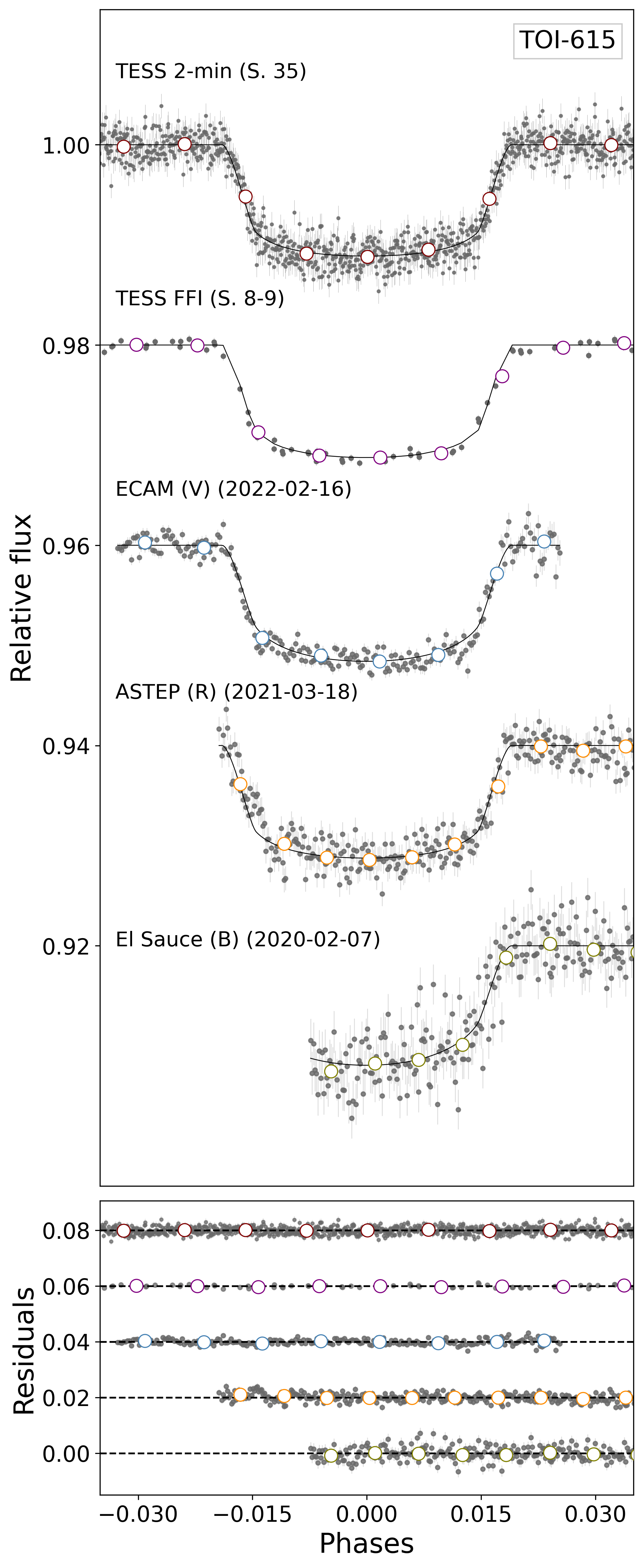}
  \hspace{0.2cm}
  \includegraphics[width=0.32\textwidth]{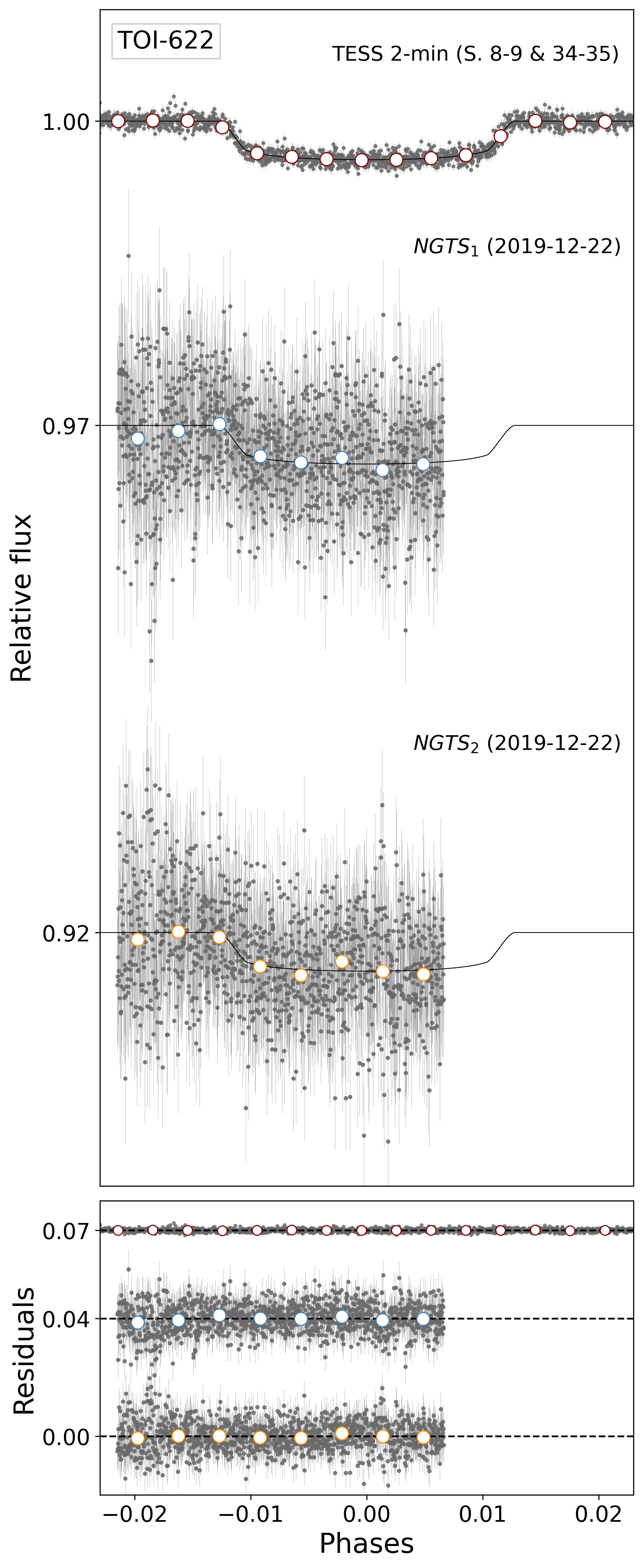}
  \includegraphics[width=0.326\textwidth]{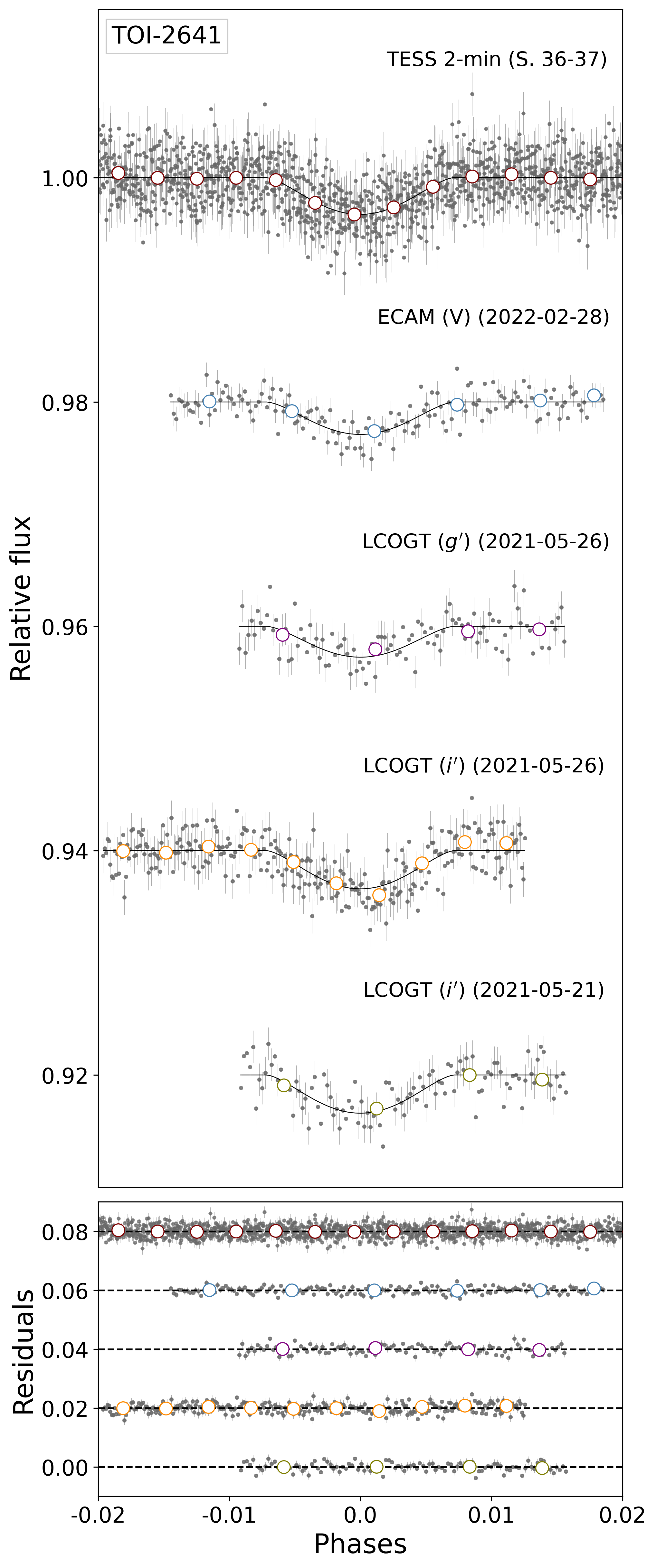}
  \caption{Phase-folded detrended TESS, EulerCam, ASTEP, EL-Sauce, NGTS, and LCOGT light curves for TOI-615b (\textit{left}), TOI-622b (\textit{middle}), and TOI-2641b (\textit{right}). The model plotted is the MCMC best-fit solution to the global model.} 
  \label{fig:Light curves}
\end{figure}

\newpage
\section{Joint analysis priors}
\input{tables/conanpriors}

\end{appendix}
\end{document}

%% file: tables/tessandgroundphotometry.tex
\begin{table}
\tiny
\caption{Summary of the TESS and ground-based photometric observations of TOI-615, TOI-622, and TOI-2641.}
\centering
\setlength{\tabcolsep}{12pt}
\begin{tabular}{lccc}
        \hline
        \hline
        Date(s) & Facility &  Detrending \\
        \hline
         \textbf{TOI-615} & &\\
         2019 Feb 02 - Mar 26 & TESS FFI & - \\
         2021 Feb 09 - Mar 07 & TESS 2-min  & GPs \\
         2020 Feb 07 & El Sauce (B) & $p(AM)$\\
         2021 Mar 18 & ASTEP ($R_c$) & -\\
         2022 Feb 16 & EulerCam (V) & $p(AM+sky^2)$\\
        \textbf{TOI-622} &&\\  
         2019 Feb 02 - Mar 26 & TESS 2-min & GPs \\
         2021 Jan 13 - Mar 07 & TESS 2-min & GPs \\
         2019 Dec 22 & NGTS$^1$ & $p(y^2+FWHM)$\\
         2019 Dec 22 & NGTS$^2$ & $p(x+y^2+FWHM)$\\
        \textbf{TOI-2641} &&\\  
         2021 Mar 07 - Apr 20 & TESS 2-min & GPs \\
         2022 Feb 28 & EulerCam (V) & $p(t+x)$\\
         2021 May 26 & LCOGT ($g'$) & $p(sky)$\\
         2021 May 26 & LCOGT ($i'$) & -\\
         2021 May 21 & LCOGT ($i'$) & $p(t)$\\
        \hline
    \end{tabular}
    \label{tab:tessandgroundphotometry}
\begin{tablenotes}
\item
\textbf{Notes:} The notation of the baseline models for, $\textit{p(j}^\textit{i}\textit{)}$, refers to a polynomial of degree \textit{i} in parameter \textit{j} (t:time, AM:airmass, FWHM:stellar FWHM and sky:sky background). The notation of GPs refers to Gaussian Processes.
\end{tablenotes}
\end{table}

%% file: tables/stellarparameters.tex
\begin{table*}
\caption{Stellar parameters of TOI-615, TOI-622, and TOI-2641.}
\centering
\renewcommand{\arraystretch}{1.1}
\begin{tabular}{lcccc}
\hline\hline
Property & TOI-615& TOI-622 & TOI-2641 & Reference \\
    \hline
    Spectral type & F2V & F6V & F9V & \citet{PecautandMmamajek2013}
    \\[4pt]
    Identifiers& 
    \\[4pt]
    TIC ID  & 190496853 & 83092282 & 162802770 & TICv8\\
    2MASS ID & J08533810-4032376 & J08211322-4629031 & J11243133-4243555 & 2MASS\\
    Gaia ID & 5524780333100885888 & 5516690878153930240 & 5382970374226953600 & $Gaia$ DR3
    \\[4pt]
    Astrometric parameters 
    \\[4pt]
    R.A. (J2000) & 08:53:38.11 & 08:21:13.22 & 11:24:31.34 & $Gaia$ DR3\\
    Dec (J2000) & -40:32:37.72 & -46:29:03.20 & -42:43:55.55 & $Gaia$ DR3\\
    Parallax (mas) & 2.82 $\pm$ 0.01 & 	8.15 $\pm$ 0.01 & 2.88 $\pm$ 0.02 & $Gaia$ DR3\\
    Distance (pc) & 354.19 $\pm$ 1.71 & 122.69 $\pm$ 0.19 & 346.79 $\pm$ 2.08 & $Gaia$ DR3\\
    $\mu_{\rm{R.A.}}$ (mas yr$^{-1}$) & -17.285 $\pm$ 0.014 &  -4.618 $\pm$ 0.013 & -32.674 $\pm$ 0.012 & $Gaia$ DR3\\
    $\mu_{\rm{Dec}}$ (mas yr$^{-1}$) & 25.695 $\pm$ 0.013 & -17.407 $\pm$ 0.014 & -9.692 $\pm$ 0.014 & $Gaia$ DR3
    \\[4pt]
    Photometric parameters 
    \\[4pt]
    TESS (mag) & 10.302 $\pm$ 0.006 & 8.562 $\pm$ 0.006 & 11.1897 $\pm$ 0.006 & TICv8\\
    B (mag) & 11.11 $\pm$ 0.05 & 9.47 $\pm$ 0.03 & 12.02 $\pm$ 0.11 & Tycho-2\\
    V (mag) & 10.82 $\pm$ 0.06 & 8.995 $\pm$ 0.002 & 11.69 $\pm$ 0.13 & Tycho-2\\
    G (mag) & 10.579 $\pm$ 0.003 & 8.8910 $\pm$ 0.0003 & 11.565 $\pm$ 0.003 & $Gaia$ DR3\\
    J (mag) & 9.88 $\pm$ 0.02 & 8.11 $\pm$ 0.02 & 10.65 $\pm$ 0.03 & 2MASS\\
    H (mag) & 9.74 $\pm$ 0.03 &7.93 $\pm$ 0.05 & 10.37 $\pm$ 0.02 & 2MASS\\
    K (mag) & 9.67 $\pm$ 0.02 & 7.86 $\pm$ 0.02 & 10.33 $\pm$ 0.02 & 2MASS
    \\[4pt]
    Bulk parameters 
    \\[4pt]
    $A_V$     &  $0.15 \pm 0.05$   & $0.05 \pm 0.05$   &  $0.11 \pm 0.08$ & Sect. \ref{sec:stellaranalysis}\\
    $F_{\rm bol}$  (10$^{-9}$ erg s$^{-1}$ cm$^{-2}$)  &  $1.519 \pm 0.035$    &  $6.44 \pm 0.15$   &  $0.593 \pm 0.028$  & Sect. \ref{sec:stellaranalysis}\\
    $L_{\rm bol}$  (L$_\odot$)  &   $5.86 \pm 0.14$   &  $2.979 \pm 0.070$     &  $2.19 \pm 0.10$ & Sect. \ref{sec:stellaranalysis}\\
    $R_\star$  (R$_\odot$)      &  $1.732 \pm 0.055$    &  $1.415 \pm 0.047$   &  $1.336 \pm 0.055$& Sect. \ref{sec:stellaranalysis}\\
    $M_\star$  (M$_\odot$)      &  $1.449 \pm 0.087$    &   $1.313 \pm 0.079$   &   $1.16 \pm 0.07$ & Sect. \ref{sec:stellaranalysis}\\
    \teff\,(K) & 6850 $\pm$ 100 & 6400 $\pm$ 100 & 6100 $\pm$ 100 & Sect. \ref{sec:stellaranalysis}\\
    \feh (dex) & -0.10 $\pm$ 0.07 & 0.09 $\pm$ 0.07 & -0.15 $\pm$ 0.08 & Sect. \ref{sec:stellaranalysis}\\
    $\log g_*$ (cm\,s$^{-2}$) & 4.2 $\pm$ 0.2 & 4.2 $\pm$ 0.2 & 4.2 $\pm$ 0.1 & Sect. \ref{sec:stellaranalysis} \\
	$v \sin i_{*}$ (km\,s$^{-1}$) & 16.3 $\pm$ 0.8 & 19.0 $\pm$ 0.9  & 1.9 $\pm$ 1.2 & Sect. \ref{sec:stellaranalysis}\\
	$P_{\rm rot}$ / $\sin i_{*}$ (days) & 5.38 $\pm$ 0.31 & 3.77 $\pm$ 0.22 & 35.6 $\pm$ 13.1 &Sect. \ref{sec:stellaranalysis} \\
	Age	(Gyr) & 1.7 $\pm$ 0.3 & 0.9 $\pm$ 0.2 & 10.8 $\pm$ 9.0 & Sect. \ref{sec:stellaranalysis}\\
    \hline
\end{tabular}
\\
\begin{tablenotes}
\item Sources: TICv8 \citep{Stassun2019}, 2MASS \citep{Skrutskie2006}, $Gaia$ DR3 \citep{GAIADR3}, Tycho-2 \citep{Tycho}
\\\textbf{}
\textbf{Notes:} The spectral type is based on the \teff~from the stellar analysis (see Section \ref{sec:stellaranalysis} and Table 5 in \citealt{PecautandMmamajek2013}). 
\end{tablenotes}
\label{tab:Stellar parameters}
\end{table*}

%% file: tables/planetparamconan.tex
\begin{table*}[h]
\caption{Maximum values and 68$\%$ confidence intervals of the posterior distributions from joint modeling with \CONAN.}
\centering
\renewcommand{\arraystretch}{1.17}
\setlength{\tabcolsep}{18pt}
\begin{center}
\begin{tabular}{lcccc}
\hline\hline
\textbf{Parameters} & \textbf{TOI-615}& \textbf{TOI-622} & \textbf{TOI-2641} \\
 \hline
    \textbf{Jump parameters} &  \\
    $\it{P}$ (days)\dotfill  & 4.6615983$^{+0.0000025}_{-0.0000016}$ & 6.402513$^{+0.000031}_{-0.000054}$& 4.880974$^{+0.000023}_{-0.000037}$\\
    $T_0$ (\bjdtdb)\dotfill & 2459259.49383$^{+0.00017}_{-0.00012}$ & 2458520.69176$^{+0.00031}_{-0.00046}$& 2459332.21332$^{+0.00064}_{-0.00067}$\\
    $R_{P}$/\rstar\dotfill & 0.10023$^{+0.0007}_{-0.00054}$  & 0.05989$^{+0.00049}_{-0.00052}$& 0.124$^{+0.031}_{-0.053}$\\
    $\textit{b}$\dotfill  & 0.476$^{+0.029}_{-0.068}$ &  0.629$^{+0.066}_{-0.091}$& 1.043$^{+0.048}_{-0.065}$\\
    $T_{14}$ (days)\dotfill& 0.1767$^{+0.0011}_{-0.0011}$ & 0.1628$^{+0.0015}_{-0.0012}$& 0.0699$^{+0.0026}_{-0.0020}$\\
    $\sqrt{e} \sin \omega$\dotfill  & 0 & 0 & 0\\
    $\sqrt{e} \cos \omega$\dotfill  & 0 & 0 & 0\\       
    $\omega_*$ (deg)\dotfill  & ... & ... & ...\\
    $\textit{K}$ (m$s^{-1}$)\dotfill  & 42.4$^{+6.9}_{-8.6}$ & 27.8$^{+6.4}_{-7.4}$& 41.6$^{+2.6}_{-3.7}$\\
    $u_{1,2min}$\dotfill  & 0.209$^{+0.030}_{-0.025}$ & 0.226$^{+0.027}_{-0.026}$ & 0.274$^{+0.028}_{-0.031}$\\
    $u_{2,2min}$\dotfill  & 0.349$^{+0.032}_{-0.040}$ & 0.287$^{+0.040}_{-0.031}$ &0.322 $^{+0.042}_{-0.041}$\\
    $u_{1,FFI}$\dotfill  & 0.209$^{+0.030}_{-0.025}$ & ... & ...\\
    $u_{2,FFI}$\dotfill  & 0.349$^{+0.032}_{-0.040}$ & ... & ...\\
    \textbf{Derived parameters} &  \\
    $\it{R_{P}}$ (\rjup)\dotfill  & 1.693$^{+0.052}_{-0.057}$  & 0.824$^{+0.028}_{-0.029}$ &1.615 $^{+0.462}_{-0.640}$\\
    $\it{M_{P}}$ (\mjup)\dotfill  & 0.435$^{+0.086}_{-0.082}$ & 0.303$^{+0.069}_{-0.072}$ & 0.386$^{+0.022}_{-0.036}$\\
    $\rho_P$ (\densjup)\dotfill  & 0.084$^{+0.018}_{-0.018}$ & 0.507$^{+0.126}_{-0.126}$ & 0.092$^{+0.078}_{-0.109}$\\
    $\rho$ (\densstar)\dotfill  & 0.276$^{+0.033}_{-0.030}$ & 0.452$^{+0.060}_{-0.047}$ & 0.482$^{+0.059}_{-0.060}$\\
    $e$ & 0 (adopted, 3$\sigma$ $<$ 0.39) & 0 (adopted, 3$\sigma$ $<$ 0.42)  & 0 (adopted, 3$\sigma$ $<$ 0.18) \\
    $i$ (degrees)\dotfill  & 86.73$^{+0.60}_{-0.22}$ & 86.62$^{+0.77}_{-0.54}$ & 83.75$^{+0.71}_{-0.55}$\\
    $\it{a}$ (AU)\dotfill  & 0.0678$^{+0.0031}_{-0.0026}$ & 0.0708$^{+0.0052}_{-0.0059}$ & 0.0607$^{+0.0042}_{-0.0043}$\\
    $\it{a/\rstar}$\dotfill  & 8.458$^{+0.214}_{-0.235}$ & 10.637$^{+0.846}_{-0.671}$ & 9.677$^{+0.599}_{-0.481}$\\
    $T_{eq}$ (K)\dotfill  & 1666$^{+24}_{-24}$& 1388$^{+22}_{-22}$& 1387$^{+22}_{-23}$\\
    $H$ (km)\dotfill  & 1601$^{+332}_{-321}$& 454$^{+108}_{-113}$& 1368$^{+786}_{-1092}$ \\
    $\fave$ (10$^{9}$ erg s$^{-1}$ cm$^{-2}$)\dotfill  & 1.75$^{+0.13}_{-0.21}$ & 0.82$^{+0.14}_{-0.16}$& 0.82$^{+0.14}_{-0.15}$ \\
    \textbf{Instrumental parameters} & \\
    $\log(\sigma_{w,2min}$)\dotfill  & -10.69$^{+0.48}_{-1.22}$ & -8.65$^{+0.03}_{-0.04}$ & -11.53$^{+1.27}_{-0.34}$\\ 
    $\log(\sigma_{GP,2min}$)\dotfill  & -7.90$^{+0.10}_{-0.18}$ & -8.65 $^{+0.08}_{-0.07}$ & -8.84$^{+0.17}_{-0.13}$ \\
    $\log(\rho_{GP,2min}$)\dotfill   & -0.57 $^{+0.12}_{-0.29}$ & -1.05$^{+0.13}_{-0.10}$ &  -0.90$^{+0.35}_{-0.43}$ \\
    $\log(\sigma_{w,FFI}$)\dotfill  & -11.76$^{+1.26}_{-0.21}$& ... & ...\\ 
    $\log(\sigma_{w,ECAM}$)\dotfill  & -7.32$^{+0.05}_{-0.13}$& ... & -7.83$^{+0.44}_{-2.61}$\\ 
    $\log(\sigma_{w,NGTS_{1}}$)\dotfill  & ... & -8.59$^{+0.41}_{-0.20}$ & ...\\ 
    $\log(\sigma_{w,NGTS_{2}}$)\dotfill  & ... & -10.42$^{+0.44}_{-0.18}$ & ...\\ 
    $\log(\sigma_{w,ASTEP}$)\dotfill  & -7.13$^{+0.11}_{-0.21}$& ... & ...\\ 
    $\log(\sigma_{w,LCO, g'}$)\dotfill  &  ... & ... & -11.67$^{+3.77}_{-0.15}$ \\
    $\log(\sigma_{w,LCO_{1} (i')}$)\dotfill  & ... & ... & -11.70$^{+3.08}_{-0.15}$ \\
    $\log(\sigma_{w,LCO_{2} (i')}$)\dotfill  & ... & ... & -8.45$^{+0.69}_{-3.11}$ \\
    $\log(\sigma_{w,ELSAUCE}$)\dotfill  & -6.46$^{+0.12}_{-0.11}$& ... & ...\\ 
    $\sigma_{w,HARPS^{1}}$ ($\kms$)\dotfill & 0.0222$^{+0.0072}_{-0.0044}$& 0.0139$^{+0.0034}_{-0.0031}$ & 0.0078$^{+0.0075}_{-0.0042}$\\ 
    $\gamma_{HARPS^{1}}$ ($\kms$)\dotfill  & 31.2805$^{+0.0044}_{-0.0081}$& 15.7987$^{+0.0052}_{-0.0056}$ & -6.5814$^{+0.0048}_{-0.0053}$\\  
    $\sigma_{w,HARPS^{2}}$ ($\kms$)\dotfill & ... & ... & 0.0134$^{+0.0094}_{-0.0063}$\\ 
    $\gamma_{HARPS^{2}}$ ($\kms$)\dotfill  & ... & ... & -6.5798$^{+0.0067}_{-0.0083}$\\  
    $\sigma_{w,CORALIE}$ ($\kms$)\dotfill & ... & ... &0.0219$^{+0.0129}_{-0.0075}$\\   
    $\gamma_{CORALIE}$ ($\kms$)\dotfill & ... & ... &-6.5865$^{+0.0071}_{-0.0081}$\\   
    $\sigma_{w,CHIRON}$ ($\kms$)\dotfill  & ... & ... & 0.036$^{+0.010}_{-0.007}$\\   
    $\gamma_{CHIRON}$ ($\kms$)\dotfill & ... & ... & -17.9880$^{+0.0098}_{-0.0063}$ \\ 
    \hline
\end{tabular}
\begin{tablenotes}
\item
\textbf{Notes:} $^{(a)}$ The description of the parameters can be found in Table \ref{tab:CONAN priors}. $^{(b)}$ The posterior estimate indicates the maximum value and then error bars the 68$\%$ credibility intervals. $^{(c)}$ The equilibrium temperature is calculated using T$_{eq}$ = \teff(1-A)$^{1/4}\sqrt{\frac{R_{*}}{2\textit{a}}}$, assuming a Bond albedo of A = 0. The scale height is calculated using $H$ = $\frac{k_{b}T_{eq}}{\mu g}$ assuming a mean molecular weight of 2.2 g/mol. The incident flux is calculated using $\fave$ = $\frac{L_{*}}{4\pi\textit{a}^{2}}$. $^{(d)}$ HARPS$^{1}$ indicates spectra collected in EGGS mode (PI: Psaridi, 108.22LR.001) and HARPS$^{2}$ indicates spectra collected in HAM mode (PI: Armstrong, 108.21YY.001).
\end{tablenotes}
\label{tab:CONAN results}
\end{center}
\end{table*}

%% file: tables/conanpriors.tex
\begin{table}[h]
\caption{Prior values for the joint photometric and radial velocity analysis of TOI-615, TOI-622, and TOI-2641 with \CONAN.}
\centering
\renewcommand{\arraystretch}{1.1}
\setlength{\tabcolsep}{4pt}
\begin{tabular}{lcccc}
\hline\hline
\textbf{Parameters} & \textbf{TOI-615}& \textbf{TOI-622} & \textbf{TOI-2641} \\
 \hline
    $\it{P}$ (days), orbital period & $\mathcal{U}(4.5, 4.8)$ &  $\mathcal{U}(6.2, 6.6)$ & $\mathcal{U}(4.6, 5.0)$ \\
    $T_0$ (\bjdtdb), time of conjunction & $\mathcal{U}(2459258, 2459260)$ & $\mathcal{U}(2458520, 2458521)$ & $\mathcal{U}(2459332, 2459333)$  \\
    $R_{P}/$\rstar, planet-to-star radius & $\mathcal{U}(0, 0.2)$ & $\mathcal{U}(0, 0.2)$ & $\mathcal{U}(0, 0.22)$\\
    $\textit{b}$, impact parameter & $\mathcal{U}(0, 1)$ & $\mathcal{U}(0, 1)$ & $\mathcal{U}(0, 1.5)$\\
    $\textit{e}$, eccentricity & $\mathcal{F}(0)$ & $\mathcal{F}(0)$ & $\mathcal{F}(0)$\\
    $\omega_*$ (deg), argument of periastron & ... & ... & ...\\
    K (m$s^{-1}$), RV semi-amplitude & $\mathcal{U}(10, 60)$ & $\mathcal{U}(10, 40)$ & $\mathcal{U}(10, 60)$ \\
    \textbf{TESS photometry:} \\
    $\log(\sigma_{w,2min}$) , jitter & $\mathcal{U}(-12,-5)$ & $\mathcal{U}(-12,-5)$ & $\mathcal{U}(-12,-5)$\\ 
    $\log(\sigma_{GP,2min}$), amplitude of the GP & $\mathcal{U}(-10,5)$ & $\mathcal{U}(-10,5)$ & $\mathcal{U}(-10,5)$\\
    $\log(\rho_{GP,2min}$), time/length-scale of the GP & $\mathcal{U}(-3,3)$ & $\mathcal{U}(-3,3)$ & $\mathcal{U}(-5,5)$\\
    $\log(\sigma_{w,FFI}$) , jitter & $\mathcal{U}(-12,-5)$ & ... & ...\\ 
    $u_{1,2min}$, limb-darkening coefficient & $\mathcal{N}(0.201,0.040)$ & $\mathcal{N}(0.239, 0.034)$ & $\mathcal{N}(0.257, 0.034)$\\
    $u_{2,2min}$, limb-darkening coefficient & $\mathcal{N}(0.321,0.044)$ & $\mathcal{N}(0.314, 0.041)$ & $\mathcal{N}(0.298, 0.039)$\\ 
    $u_{1,FFI}$, limb-darkening coefficient & $\mathcal{N}(0.201,0.040)$ & ... & ...\\
    $u_{2,FFI}$, limb-darkening coefficient & $\mathcal{N}(0.321,0.044)$ & ... & ...\\ 
    \textbf{Ground based photometry:} \\
    $u_{1,ECAM}$, limb-darkening coefficient& $\mathcal{F}(0.379)$ & ... & $\mathcal{F}(0.443)$\\
    $u_{2,ECAM}$, limb-darkening coefficient & $\mathcal{F}(0.311)$ &  ... & $\mathcal{F}(0.286)$\\    
    $\log(\sigma_{w,ECAM}$) , jitter & $\mathcal{U}(-12,-5)$ &... & $\mathcal{U}(-12,-5)$\\ 
    $u_{1,NGTS_{1,2}}$, limb-darkening coefficient & ... &  $\mathcal{F}(0.289)$ & ...\\
    $u_{2,NGTS_{1,2}}$, limb-darkening coefficient & ... &  $\mathcal{F}(0.316)$ & ...\\  
    $\log(\sigma_{w,NGTS_{1,2}}$) , jitter & ... & $\mathcal{U}(-12,-5)$ & ...\\ 
    $u_{1,ASTEP}$, limb-darkening coefficient & $\mathcal{F}(0.251)$ & ... & ...\\
    $u_{2,ASTEP}$, limb-darkening coefficient & $\mathcal{F}(0.331)$ & ... & ...\\   
    $\log(\sigma_{w,ASTEP}$) , jitter & $\mathcal{U}(-12,-5)$ & ...& ...\\
    $u_{1,LCO, g'}$, limb-darkening coefficient & ... & ... & $\mathcal{F}(0.547)$\\
    $u_{2,LCO, g'}$, limb-darkening coefficient & ... &  ... & $\mathcal{F}(0.217)$\\   
    $\log(\sigma_{w,LCO, g'}$) , jitter & ...& ...& $\mathcal{U}(-12,-5)$\\ 
    $u_{1,LCO_{1,2} (i')}$, limb-darkening coefficient & ... & ... & $\mathcal{F}(0.259)$\\
    $u_{2,LCO_{1,2} (i')}$, limb-darkening coefficient & ... & ... & $\mathcal{F}(0.298)$\\
    $\log(\sigma_{w,LCO_{1,2}, i'}$) , jitter & ... & ...& $\mathcal{U}(-12,-5)$\\ 
    $u_{1,ELSAUCE}$, limb-darkening coefficient & $\mathcal{F}(0.511)$ & ... & ...\\
    $u_{2,ELSAUCE}$, limb-darkening coefficient & $\mathcal{F}(0.269)$ & ... & ...\\   
    $\log(\sigma_{w,ELSAUCE}$) , jitter & $\mathcal{U}(-12,-5)$ & ... & ...\\
    \textbf{Radial velocities:} \\
    $\gamma_{HARPS^{1}}$ ($\kms$), systemic RV &$\mathcal{U}(30, 33)$ & $\mathcal{U}(14, 16)$
    &$\mathcal{U}(-8, -5)$\\  
    $\sigma_{w,HARPS^{1}}$ ($\kms$), jitter &$\mathcal{U}(0, 100)$ & $\mathcal{U}(0, 100)$ &$\mathcal{U}(0, 100)$\\
    $\gamma_{HARPS^{2}}$ ($\kms$), systemic RV & ... & ... &$\mathcal{U}(-8,-5)$\\  
    $\sigma_{w,HARPS^{2}}$ ($\kms$), jitter &... & ... & $\mathcal{U}(0, 100)$\\
    $\gamma_{CORALIE}$ ($\kms$), systemic RV & ... & ... & $\mathcal{U}(-8, -5)$ \\ 
    $\sigma_{w,CORALIE}$ ($\kms$), jitter & ... & ... & $\mathcal{U}(0, 100)$ \\
    $\gamma_{CHIRON}$ ($\kms$), systemic RV & ... & ... &$\mathcal{U}(-19,-16)$ \\    
    $\sigma_{w,CHIRON}$ ($\kms$), jitter & ... & ... &$\mathcal{U}(0, 100)$ \\

    \hline
\end{tabular}
\\
\begin{tablenotes}
            \item\textbf{Notes:} $^{(a)}$ For the priors, $\mathcal{N}$($\mu$, $\sigma^{2}$) indicates a normal distribution with mean $\mu$ and variance $\sigma^{2}$, $\mathcal{U}$(\textit{a}, \textit{b}) a uniform distribution between \textit{a} and \textit{b}, log$\mathcal{U}$(\textit{a}, \textit{b}) a log-uniform distribution between \textit{a} and \textit{b} and $\mathcal{F}$(\textit{a}) a parameter fixed to value \textit{a}.\\
            $^{(b)}$ HARPS$^{1}$ indicates spectra collected in EGGS mode (PI: Psaridi, 108.22LR.001) and HARPS$^{2}$ indicates spectra collected in HAM mode (PI: Armstrong, 108.21YY.001).
        \end{tablenotes}
\label{tab:CONAN priors}
\end{table}